\documentclass[10pt,aps,prd,twocolumn,showpacs,superscriptaddress,eqsecnum,longbibliography,nofootinbib]{revtex4-2}
\usepackage{mathrsfs,amsmath,amsthm,latexsym,amssymb,amsfonts,epsfig,cancel,enumerate,graphicx,subcaption,txfonts,overpic}
\usepackage{multirow}
\DeclareCaptionJustification{flushlast}{\leftskip=0pt \rightskip=0pt \parfillskip=0pt plus 1fil}% [rev] justified caption with left-flush last line
\usepackage{xcolor}
\definecolor{navy}{RGB}{0,0,150}
\usepackage[colorlinks, linkcolor=blue, citecolor=blue, urlcolor=blue, plainpages=false, pdfstartview=FitH]{hyperref}
\usepackage{appendix}
\allowdisplaybreaks
\begin{document}
	
	\title{Observational Signatures of Rotating Ay\'on-Beato--Garc\'{\i}a Black Holes: Shadows, Accretion Disks and Images}
	
	\author{Zhenglong Ban}
	\email{zlban123@163.com}
	\affiliation{College of Physics Science and Technology, Hebei University, Baoding 071002, China}
	
	\author{Meng Chen}
	\email{17828812028@163.com}
	\affiliation{College of Physics Science and Technology, Hebei University, Baoding 071002, China}
	
	\author{Rong-Jia Yang \footnote{Corresponding author}}
	\email{yangrongjia@tsinghua.org.cn}
	\affiliation{College of Physics Science and Technology, Hebei University, Baoding 071002, China}
	\affiliation{Hebei Key Lab of Optic-Electronic Information and Materials, Hebei University, Baoding 071002, China}
	\affiliation{National-Local Joint Engineering Laboratory of New Energy Photoelectric Devices, Hebei University, Baoding 071002, China}
	\affiliation{Key Laboratory of High-precision Computation and Application of Quantum Field Theory of Hebei Province, Hebei University, Baoding 071002, China}
	
	\begin{abstract}
We investigate the shadows, accretion disks, and observational images of rotating Ay\'on-Beato--Garc\'{\i}a (ABG) black holes with mass $M$, spin $a$, and nonlinear-electrodynamics (NLED) charge parameter $\zeta$, treating photons as neutral test particles on null geodesics of the background metric so as to obtain the purely geometric shadow. The shadow shrinks with increasing $\zeta$ and develops a ``D''-shaped morphology for near-extremal spin. The thermal disk properties follow from the Novikov--Thorne model with inner edge at the innermost stable circular orbit (ISCO), whereas the images use a separate, phenomenological optically thin emission model extending to the horizon. The image asymmetry and redshift maps depend strongly on $(a, \zeta)$ and the inclination. Comparing the geometric shadow diameters with Event Horizon Telescope observations of M87$^{*}$ and Sgr A$^{*}$ gives the indicative estimates $\zeta \lesssim 0.21\,M$ and $\zeta \lesssim 0.43\,M$, respectively; since $\zeta$ characterizes each black hole individually, these are quoted separately rather than combined, and the Kerr limit $\zeta = 0$ remains fully consistent with both.
\end{abstract}
	
	\maketitle
	\section{Introduction}
	Black holes (BHs), predicted by Einstein's theory of General Relativity (GR), represent one of the most intriguing consequences of modern theoretical physics—regions where gravity overwhelms all other forces. Gravitational waves emitted during black hole mergers have been successfully detected by the Laser Interferometer Gravitational-Wave Observatory (LIGO), providing a landmark confirmation of General Relativity and establishing a powerful new method for investigating black hole properties \cite{LIGOScientific:2016aoc}. In a complementary breakthrough, the Event Horizon Telescope (EHT) Collaboration achieved the first visual confirmation of a BH in 2019 by imaging the shadow of M87 $ ^{*} $ \cite{EventHorizonTelescope:2019dse,EventHorizonTelescope:2019uob,EventHorizonTelescope:2019jan,EventHorizonTelescope:2019ths,EventHorizonTelescope:2019pgp,EventHorizonTelescope:2019ggy}, and subsequently captured the image of Sgr A $ ^{*}$, the supermassive BH at the Milky Way's center \cite{EventHorizonTelescope:2022wkp,EventHorizonTelescope:2022apq,EventHorizonTelescope:2022wok,EventHorizonTelescope:2022exc,EventHorizonTelescope:2022urf,EventHorizonTelescope:2022xqj}. These observations provide critical empirical anchors for theoretical models and carry imprints of the complex physical processes occurring in the immediate vicinity of event horizons.
	
	Motivated by the EHT's imaging of M87 $ ^{*}$ and Sgr A $ ^{*}$, the black hole shadow has become a central tool for probing the nature of compact objects. This dark region in the sky arises from the absence of light rays that fall into the event horizon, with its contour determined by the spacetime's photon sphere. The foundational study by Synge \cite{Synge:1966okc} showed that a Schwarzschild BH casts a perfectly circular shadow, a result refined by Luminet \cite{Luminet:1979nyg} through radiative transfer calculations. Bardeen \cite{Bardeen:1973tla} then demonstrated that rotation breaks this symmetry, yielding an oblate shadow whose deformation encodes the BH's spin and inclination. In more general settings, spherically symmetric BHs (e.g., charged or regular) retain circular shadows whose radii reflect departures from GR \cite{Roy:2020dyy,Guo:2021wid,Heydari-Fard:2021pjc,dePaula:2023ozi}, while rotating solutions in alternative theories produce distinct asymmetries \cite{Abdujabbarov:2016hnw,Amir:2016cen,Sharif:2016znp,Wei:2013kza,Abdujabbarov:2012bn,Neves:2020doc,Amarilla:2013sj,Atamurotov:2013sca,Mishra:2019trb,Sarikulov:2022atq}. Consequently, shadow measurements now serve not only to verify the Kerr paradigm but also to constrain dark matter profiles, extra dimensions, and modified gravity \cite{Papnoi:2014aaa,Abdujabbarov:2015rqa,Konoplya:2019sns,Jusufi:2020odz,Sau:2022afl}. In \cite{Ban:2024qsa}, we constructed two rotating quantum-corrected black holes in effective loop quantum gravity, investigated their shadow properties, and showed that the quantum parameter can be constrained by EHT observations of M87$ ^{*}$  and Sgr A$^{*}$.
	
	Beyond the shadow's boundary, the structure and emission of the accretion disk serve as a sensitive probe of the innermost spacetime geometry. Black hole's accretion is a perennially significant research topic. From an astrophysical perspective, it is fundamental to understanding astronomical phenomena such as active galactic nuclei, X-ray binaries, and gamma-ray bursts \cite{Yuan:2014gma}. In the canonical picture, diffuse material—gas and dust—loses angular momentum and spirals toward the central object, converting gravitational binding energy into electromagnetic radiation. The foundational thin-disk model of Shakura and Sunyaev was later elevated to a fully general-relativistic treatment by Novikov, Thorne, and Page \cite{Shakura:1972te,Page:1974he}, who accounted for frame-dragging, redshift, and the location of the innermost stable circular orbit (ISCO). This formalism has been extensively generalized to black holes in alternative theories of gravity, regular spacetimes, and charged metrics, where deviations from Kerr manifest in altered flux profiles and spectral shapes \cite{Harko:2009rp,Chen:2011wb,Liu:2021yev,Heydari-Fard:2021ljh,Karimov:2018whx,Chen:2011rx,Kazempour:2022asl,Gyulchev:2019tvk,Wu:2024sng,Feng:2024iqj,Liu:2024brf,Yin:2025coq,Yang:2024lmj}. As shown in  \cite{Ban:2026tyh}, the Hernquist dark matter (DM) halo modifies the ISCO and marginally bound orbit (MBO) radii, leaving distinct imprints on both gravitational waveforms and disk images. As a result, the disk's thermal spectrum not only reveals the mass and spin of the compact object but also encodes subtle signatures of new physics, making it a vital tool for testing General Relativity in the strong-field regime \cite{Bambi:2015kza}.
	
	Beyond spectral signatures, the spatial morphology of black hole images—shaped by gravitational lensing of accretion emission—offers another critical test of strong-field gravity. While spherically symmetric accretion yields a simple shadow \cite{Narayan:2019imo}, the inclusion of a geometrically thin accretion disk reveals a bright ring structure surrounding the shadow, as demonstrated in the context of Schwarzschild spacetime \cite{Gralla:2019xty}. This framework has since been extended to static black holes and wormhole geometries in a variety of modified gravity theories \cite{Konoplya:2021slg,Chowdhuri:2020ipb,Tsukamoto:2014tja,Cunha:2015yba,He:2022yse,Zhang:2024jrw,Gan:2021xdl}. For rotating spacetimes, the interplay of spin, external magnetic fields, and observer inclination has been shown to produce complex, multiringed images \cite{Hou:2022eev}. Similar analyses in alternative rotating black hole backgrounds further confirm that deviations from the Kerr metric manifest as characteristic distortions in the observed image, providing robust diagnostics for testing General Relativity \cite{Meng:2025ivb}.
	
	A compelling strategy for resolving spacetime singularities while preserving physical energy conditions is the coupling of gravity to nonlinear electrodynamics (NLED). First proposed by Born and Infeld to eliminate the infinite self-energy of point charges \cite{Born:1934gh}, NLED has since evolved into a versatile framework with deep connections to cosmology \cite{DeLorenci:2002mi,Novello:2003kh,Mignani:2016fwz} and string theory \cite{Seiberg:1999vs,Fradkin:1985qd}. Within this context, the Ay\'on-Beato--Garc\'{\i}a (ABG) black hole emerged as a seminal example of a regular, electrically charged solution satisfying the weak energy condition \cite{Ayon-Beato:1998hmi}. In our previous work, we investigated the shadow and thin accretion disk around the static ABG black hole coupled with a cloud of strings \cite{Cai:2025pan}. Its rotating counterpart was later constructed using the modified Newman–Janis algorithm \cite{Azreg-Ainou:2014pra}. In this paper, we compute the shadow silhouette, thermal spectrum from a Novikov–Thorne accretion disk, and synthetic observational images of this rotating ABG spacetime. By comparing these predictions with the Event Horizon Telescope measurements of M87$^{*}$ and Sgr A$^{*}$, we obtain indicative estimates of the NLED charge parameter within a purely geometric treatment of photon propagation; this assumption is discussed in Sec.~\ref{section2}. Compared with previous studies of rotating regular black holes, the main new ingredients of the present work are: (i) a systematic imaging study of the rotating ABG spacetime, in which the emitting region of the optically thin disk model extends down to the event horizon and includes the plunging flow inside the ISCO; (ii) redshift maps of both the direct and the lensed images for prograde and retrograde flows; and (iii) a shadow-based comparison with both M87$^{*}$ and Sgr A$^{*}$, presented together with a critical assessment of the underlying assumptions and of the resulting bounds.
	
	The structure of the paper is as follows. The rotating Ay\'on-Beato--Garc\'{\i}a black hole, along with its event horizon and shadow, is analyzed in Sec.~\ref{section2}. The radiative properties of thin accretion disks—including energy flux, radiation temperature, and spectral energy distribution—are examined in Sec.~\ref{section3}. Observational images of the black hole illuminated by a thin accretion disk are computed in Sec.~\ref{section4}. Constraints on the model parameters are derived in Sec.~\ref{section5} through comparison of theoretical shadow sizes with EHT observations of M87$^{*}$ and Sgr A$^{*}$. A summary of the results and their implications is provided in Sec.~\ref{section6}.
	
	\section{The Rotating Ay\'on-Beato--Garc\'{\i}a Black Hole: Event Horizon and Shadow}
	\label{section2}
	In this section, we present the rotating Ay\'on-Beato--Garc\'{\i}a black hole solution and analyze its event horizon structure and shadow properties. The spacetime metric, expressed in Boyer–Lindquist coordinates, is given by \cite{Azreg-Ainou:2014pra}:
	\begin{equation}
		\begin{split}
			\mathrm{d}s^2 &= -\left(1-\frac{2f}{\rho^{2}}\right)\mathrm{d} t^{2} + \frac{\rho^{2}}{\Delta}\mathrm{d} r^{2} + \rho^{2}\mathrm{d} \theta^{2} + \frac{\Sigma\sin^{2}\theta}{\rho^{2}}\mathrm{d}\phi^{2}\\
			&\quad - \frac{4af\sin^{2}\theta}{\rho^{2}}\mathrm{d}t\mathrm{d}\phi,\label{xianyuan}
		\end{split}
	\end{equation}
	where 
	\begin{equation}
		\begin{split}
			&f=\frac{r^{2}(1-F)}{2},\\
			&\rho^{2}=r^{2}+a^{2}\cos^{2}\theta,\\
			&\Delta=r^{2}F + a^{2},\\
			&\Sigma=\left(a^{2}+r^{2}\right)^2-a^2 \Delta\sin ^2\theta,\\
			&F=1-\frac{2Mr^2}{(r^2+\zeta^2)^{3/2}}+\frac{\zeta^{2}r^2}{(r^2+\zeta^2)^{2}}.\label{dugui}
		\end{split}
	\end{equation}
	The parameters $ a $ and $ \zeta $ correspond to the black hole's spin and the NLED charge parameter (the electric charge of the seed ABG solution), respectively. The metric is seen to approach the Kerr solution in the limit $ \zeta \to 0 $, and the Schwarzschild solution is recovered when both $ \zeta \to 0 $ and $ a = 0 $.

Several remarks on the status of the metric (\ref{xianyuan}) are in order. The static ABG solution is an exact, regular solution of general relativity coupled to NLED satisfying the weak energy condition \cite{Ayon-Beato:1998hmi}. Its rotating counterpart (\ref{xianyuan}) is generated by the modified Newman--Janis algorithm \cite{Azreg-Ainou:2014pra}: the resulting metric solves the Einstein equations with an anisotropic fluid source, but an exact NLED Lagrangian sourcing the rotating configuration is not known, and rotating metrics generated in this way generally satisfy the Einstein--NLED field equations only approximately \cite{Toshmatov:2017zpr}. Furthermore, the regularity and the energy conditions of the seed solution are not automatically inherited by the rotating spacetime \cite{Bambi:2013ufa,Toshmatov:2017zpr}. We therefore regard the metric (\ref{xianyuan}) primarily as a phenomenological, singularity-free deformation of the Kerr geometry, in the same spirit as many rotating regular black hole models employed in the shadow and imaging literature \cite{Abdujabbarov:2016hnw,Amir:2016cen}. We also note that recent perturbative analyses indicate that nonsingular black holes supported by NLED are generically prone to Laplacian instabilities near the regular center \cite{DeFelice:2024seq}; if such instabilities afflict the ABG family, the geometry considered here would describe at best a long-lived transient configuration. A quantitative assessment of these issues lies beyond the scope of this paper.
	
\begin{figure}[htbp]
		\centering
		\begin{subfigure}{0.45\textwidth}
			\includegraphics[width=3.2in, height=5.5in, keepaspectratio]{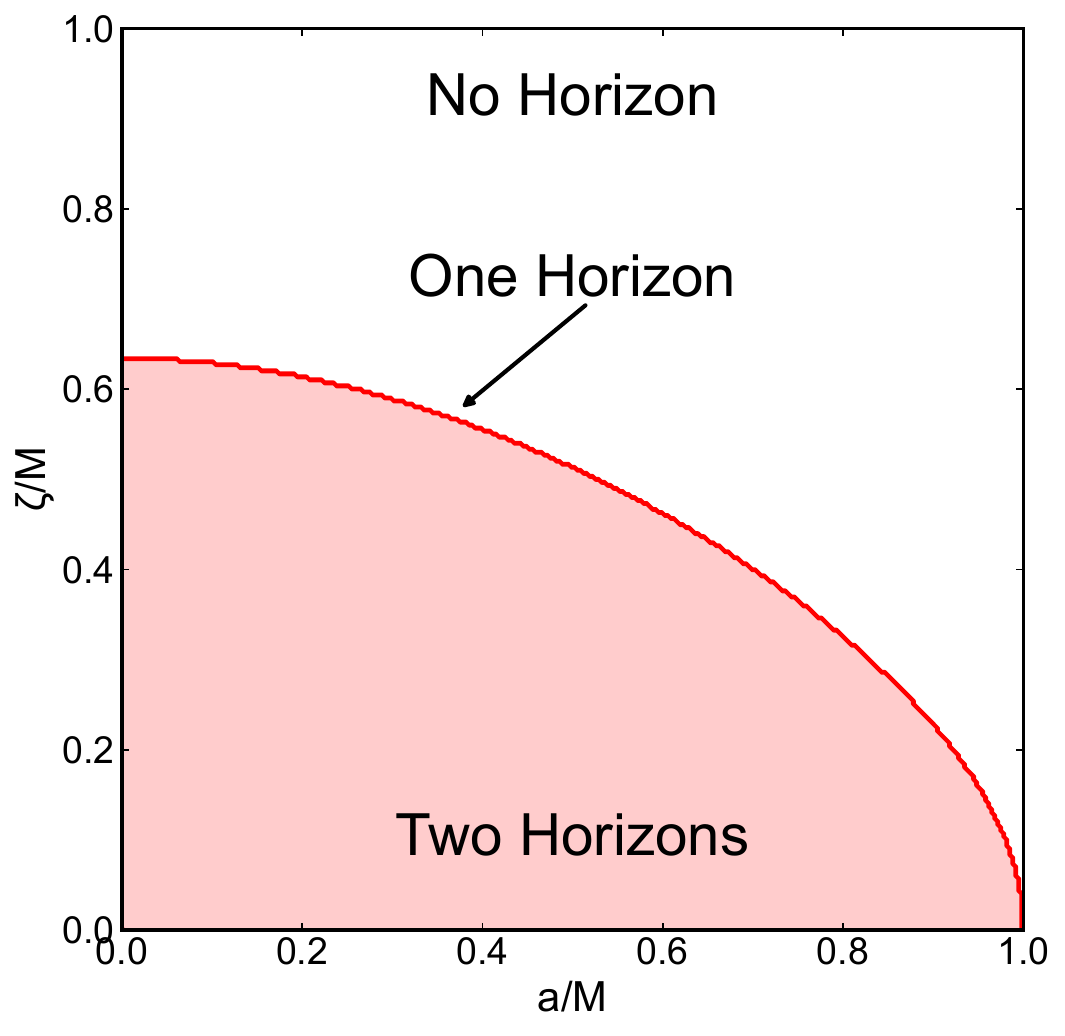}
		\end{subfigure}
		\caption{The parameter space $(a, \zeta)$ for two horizons, one degenerate horizon, and no horizon.}
		\label{horizon}
	\end{figure}

	The locations of the horizons for the rotating Ay\'on-Beato--Garc\'{\i}a black hole are obtained by solving $ g^{\mathrm{rr}} = 0 $, which leads to the following equation:
	\begin{equation}
		\Delta=r^{2}F + a^{2}=0.\label{data}
	\end{equation}
	A segment of the parameter space $ (a, \zeta) $ relevant to horizon existence is presented in Fig.~\ref{horizon}. The extremal configuration is indicated by the red solid curve, while the physically allowed black hole region is shaded in pink.
	
	Before computing the shadow, we specify our treatment of photon propagation. In NLED, electromagnetic waves do not propagate along null geodesics of the background metric but along null geodesics of an effective optical geometry, which generically exhibits birefringence \cite{Novello:1999pg}. In this work we instead follow null geodesics of the background metric (\ref{xianyuan}); that is, we compute the \emph{purely geometric shadow} of the spacetime, namely the shadow that would be probed by neutral test photons or by any field minimally coupled to the metric. This choice is motivated as follows. First, the effective optical geometry is determined by the NLED Lagrangian and by the associated electromagnetic field strength \cite{Novello:1999pg}; since no exact NLED Lagrangian and field configuration sourcing the rotating metric (\ref{xianyuan}) are known, the ingredients required for this construction are not available, and the effective geometry cannot be defined unambiguously. Second, the geometric shadow provides a well-defined benchmark that isolates the imprint of the spacetime geometry itself, and it coincides with the quantity computed in most studies of rotating regular black holes with which we compare. It should be noted that, for the actual electromagnetic radiation observed by the EHT, NLED corrections may modify the shadow size, as illustrated for a static ABG-type model in Ref. \cite{Ramadhan:2023ogm}; the bounds derived in Sec.~\ref{section5} must therefore be interpreted within this geometric framework.

The null geodesics necessary to determine the shadow of the rotating Ay\'on-Beato--Garc\'{\i}a black hole are obtained via the Hamilton–Jacobi formalism developed by Carter \cite{Carter:1968rr}. The Jacobi action $ S $ is governed by the equation
	\begin{equation}
		\frac{\partial S}{\partial \tau} = -\frac{1}{2} g^{\mathrm{\mu\nu}} \frac{\partial S}{\partial x^\mathrm{\mu}} \frac{\partial S}{\partial x^\mathrm{\nu}},
		\label{yakb}
	\end{equation}
	where $ \tau $ denotes an affine parameter. Due to the symmetries of the metric, $ S $ is assumed to take the separable form
	\begin{equation}
		S = \frac{1}{2} m^{2} \tau - E t + L \phi + S_{\mathrm{r}}(r) + S_{\mathrm{\theta}}(\theta),
		\label{action}
	\end{equation}
	with $ m = 0 $ for photons. Upon substitution into Eq. (\ref{yakb}), the geodesic equations are obtained as
	\begin{align}
		\rho^{2}\frac{\mathrm{d}t}{\mathrm{d}\tau} &= a(L - aE\sin^{2}\theta) + \frac{r^{2}+a^{2}}{\Delta}\left[(r^{2}+a^{2})E - aL\right], \label{tdian} \\
		\rho^{2}\frac{\mathrm{d}r}{\mathrm{d}\tau} &= \pm \sqrt{R(r)}, \label{rdian} \\
		\rho^{2}\frac{\mathrm{d}\theta}{\mathrm{d}\tau} &= \pm \sqrt{\Theta(\theta)}, \label{seidian} \\
		\rho^{2}\frac{\mathrm{d}\phi}{\mathrm{d}\tau} &= \frac{L}{\sin^{2}\theta} - aE + \frac{a}{\Delta}\left[(r^{2}+a^{2})E - aL\right], \label{faidian}
	\end{align}
	with the associated potentials defined by
	\begin{align}
		R(r) &= \left[(r^{2}+a^{2})E - aL\right]^{2} - \Delta \left[ C + (L - aE)^{2} \right], \label{dar} \\
		\Theta(\theta) &= C + \left(a^{2}E^{2} - L^{2}\csc^{2}\theta\right)\cos^{2}\theta. \label{dasei}
	\end{align}
	Here, $ C $ is identified as the Carter constant, ensuring the complete integrability of the geodesic system \cite{Carter:1968rr}.
	
\begin{figure*}[htbp]
		\centering
		\begin{subfigure}{0.38\textwidth}
			\includegraphics[width=\linewidth]{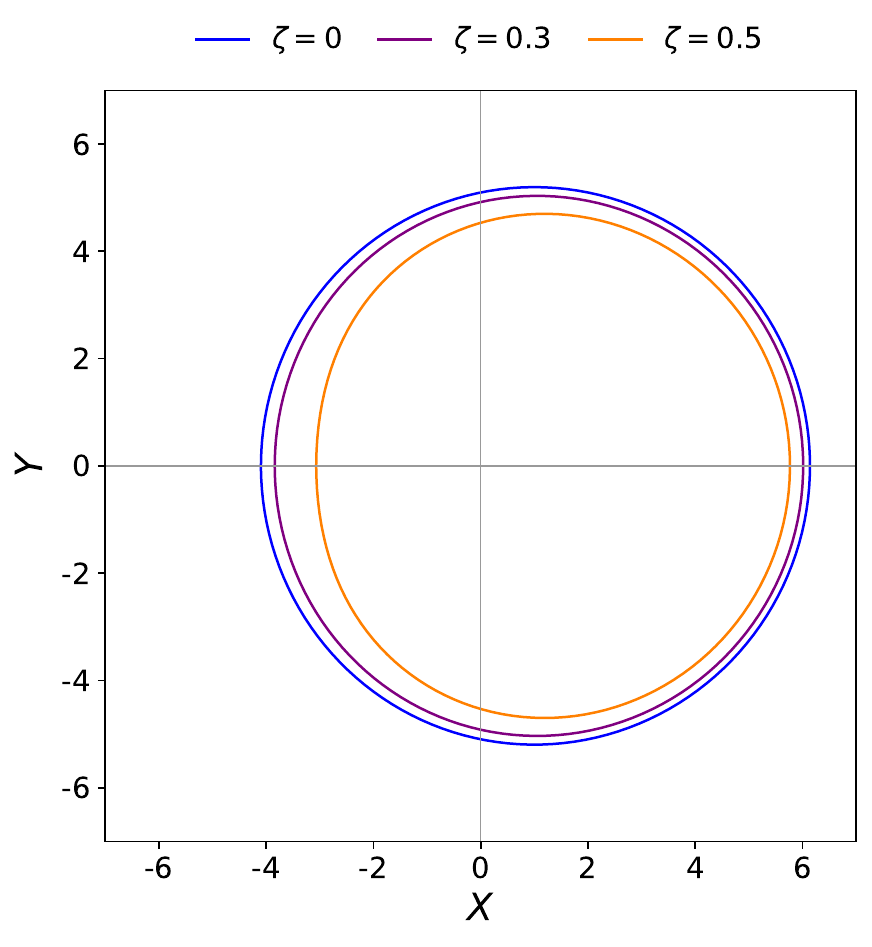}
		\end{subfigure}%
		\hspace{0.13\textwidth}% 
		\begin{subfigure}{0.38\textwidth}
			\includegraphics[width=\linewidth]{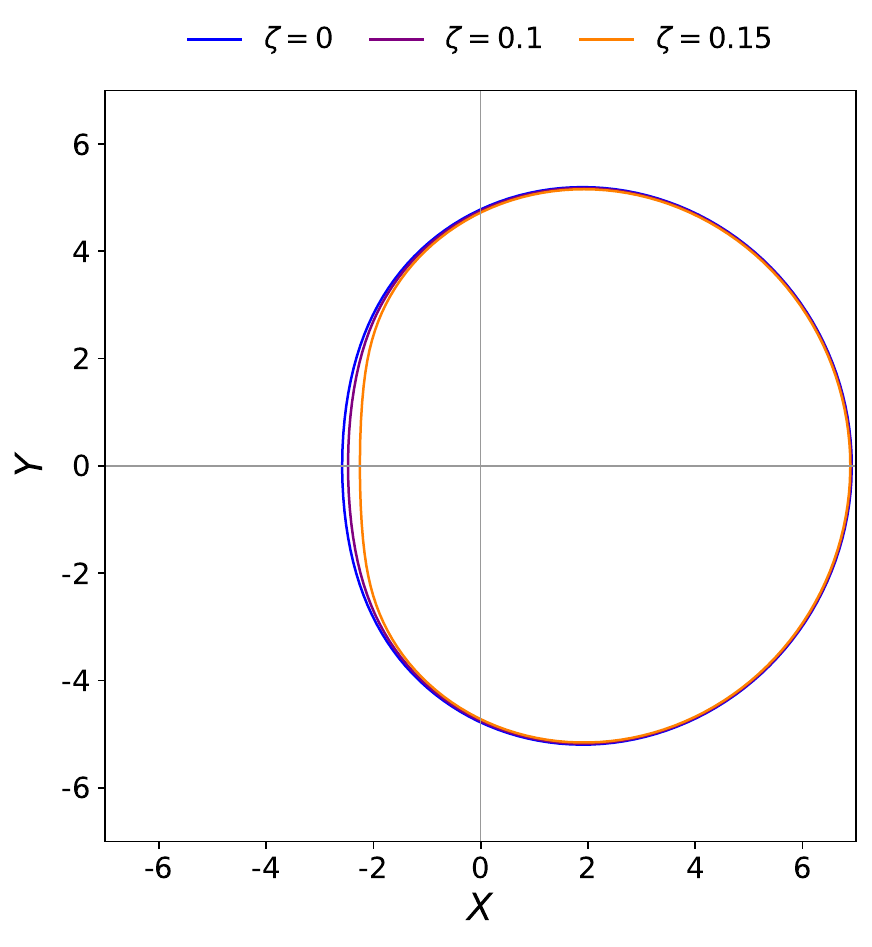}
		\end{subfigure}
		\caption{Two-dimensional shadows of the rotating Ay\'on-Beato--Garc\'{\i}a black hole for varying spin $ a $ and charge $ \zeta $. (Left) $ a = 0.5 $; (Right) $ a = 0.95 $ .}
		\label{shadow}
	\end{figure*}

	The trajectory of a photon in the rotating Ay\'on-Beato--Garc\'{\i}a spacetime is fully characterized by two dimensionless impact parameters
	\begin{equation}
		\xi = \frac{L}{E}, \quad \eta = \frac{C}{E^{2}},
		\label{kesaiyita}
	\end{equation}
	where $ E $, $ L $ and $ C $ denote the conserved energy, the axial angular momentum, and the Carter constant, respectively. To determine the boundary of the black hole shadow, one must identify the locus of unstable circular photon orbits—so-called photon spheres—which satisfy the critical conditions
	\begin{equation}
		R(r)\big|_{r=r_{\mathrm{ps}}} = 0, \quad 
		\left.\frac{\mathrm{d}R(r)}{\mathrm{d}r}\right|_{r=r_{\mathrm{ps}}} = 0, \quad 
		\left.\frac{\mathrm{d}^{2}R(r)}{\mathrm{d}r^{2}}\right|_{r=r_{\mathrm{ps}}} > 0,
		\label{tiaojian}
	\end{equation}
	with $ r_{\mathrm{ps}} $ denoting the radius of such an orbit. Solving this system yields the critical impact parameters $ (\xi, \eta) $ that define the edge of the shadow. These are explicitly given by
	\begin{align}
		\xi &= \frac{a^{2} + r_{\mathrm{ps}}^{2} - \displaystyle\frac{4\,\Delta(r_{\mathrm{ps}})\,r_{\mathrm{ps}}}{\Delta'(r_{\mathrm{ps}})}}{a}, \label{kesai} \\
		\eta &= \frac{-16\,\Delta(r_{\mathrm{ps}})^{2} r_{\mathrm{ps}}^{2} - r_{\mathrm{ps}}^{4} \Delta'(r_{\mathrm{ps}})^{2} + 8\,\Delta(r_{\mathrm{ps}}) r_{\mathrm{ps}} \left(2a^{2} r_{\mathrm{ps}} + r_{\mathrm{ps}}^{2} \Delta'(r_{\mathrm{ps}})\right)}{a^{2} \Delta'(r_{\mathrm{ps}})^{2}}. \label{yita}
	\end{align}
	Figure \ref{shadow} displays the two-dimensional silhouette of the shadow cast by a rotating Ay\'on-Beato--Garc\'{\i}a black hole for selected values of the charge parameter  $ \zeta $ . The left panel corresponds to a non-extremal configuration with spin  $ a = 0.5 $ , while the right panel shows a near-extremal case with  $ a = 0.95 $ . In both scenarios, the shadow size decreases monotonically as  $ \zeta $  increases, reflecting the weakening of gravitational lensing due to the repulsive effect of the nonlinear electromagnetic field. Notably, in the near-extremal regime ($ a = 0.95 $), the shadow undergoes a pronounced deformation—evolving from an almost circular shape at small  $ \zeta $  to a distinct ``D''-like morphology at larger values. This asymmetry arises from the interplay between rapid rotation and the charge-induced modification of the spacetime geometry, which shifts the photon capture region preferentially toward the co-rotating side and compresses the shadow on the counter-rotating side.
	
	\section{Radiative Properties of Thin Accretion Disks around Rotating Ay\'on-Beato--Garc\'{\i}a Black Holes}\label{section3}
	A brief overview of the physical properties of thin accretion disks, as formulated in the Novikov–Thorne model \cite{NT}, is first presented. The influence of the spin parameter $ a $ and charge parameter $ \zeta $ on the energy flux, radiation temperature, and observable luminosity is then examined in detail. The analysis is restricted to the equatorial ($ \theta = \pi/2 $) timelike geodesics. The corresponding Lagrangian is given by
	\begin{equation}
		\mathcal{L} = -g_{\mathrm{tt}}\dot{t}^{2} + 2g_{\mathrm{t\phi}}\dot{t}\dot{\phi} + g_{\mathrm{rr}}\dot{r}^{2} + g_{\mathrm{\phi\phi}}\dot{\phi}^{2},
		\label{la}
	\end{equation}
	with the overdot representing differentiation with respect to the proper time $ \tau $. From the Euler–Lagrange equations for $ t $  and $ \phi $, the conserved specific energy $ E $  and the angular momentum $ L $  are obtained
	\begin{equation}
		\begin{split}
			g_{\mathrm{tt}}\dot{t} + g_{\mathrm{t\phi}}\dot{\phi} &= -E, \\
			g_{\mathrm{t\phi}}\dot{t} + g_{\mathrm{\phi\phi}}\dot{\phi} &= L.
		\end{split}
		\label{EL}
	\end{equation}
	Inversion of these relations yields the four-velocity components
	\begin{equation}
		\begin{split}
			\dot{t} &= \frac{E g_{\mathrm{\phi\phi}} + L g_{\mathrm{t\phi}}}{g_{\mathrm{t\phi}}^{2} - g_{\mathrm{tt}} g_{\mathrm{\phi\phi}}}, \\
			\dot{\phi} &= -\frac{E g_{\mathrm{t\phi}} + L g_{\mathrm{tt}}}{g_{\mathrm{t\phi}}^{2} - g_{\mathrm{tt}} g_{\mathrm{\phi\phi}}}.
		\end{split}
		\label{tfai}
	\end{equation}
	The radial dynamics of timelike geodesics is constrained by the normalization condition $ g_{\mu\nu}\dot{x}^\mu \dot{x}^\nu = -1 $, which leads to the relation
	\begin{equation}
		g_{\mathrm{rr}} \dot{r}^2 = V_{\mathrm{eff}}.
		\label{Veff}
	\end{equation}
	The effective potential is given by
	\begin{equation}
		V_{\mathrm{eff}} = -1 + \frac{E^{2} g_{\mathrm{\phi\phi}} + 2 E L g_{\mathrm{t\phi}} + L^{2} g_{\mathrm{tt}}}{g_{\mathrm{t\phi}}^{2} - g_{\mathrm{tt}} g_{\mathrm{\phi\phi}}}.
		\label{Veff2}
	\end{equation}
	In the equatorial plane ($ \theta = \pi/2 $), circular orbits are defined by the conditions $ V_{\mathrm{eff}}(r) = 0 $ and $ \partial_\mathrm{r} V_{\mathrm{eff}}(r) = 0 $. These yield the angular velocity $ \Omega = \mathrm{d}\phi/\mathrm{d}t $ and the conserved quantities $ E $ and $ L $  as functions of the orbital radius
	\begin{align}
		\Omega &= \frac{-g_{\mathrm{t\phi,r}} + \sqrt{(g_{\mathrm{t\phi,r}})^{2} - g_{\mathrm{tt,r}} g_{\mathrm{\phi\phi,r}}}}{g_{\mathrm{\phi\phi,r}}}, \label{oumiga} \\
		E &= -\frac{g_{\mathrm{tt}} + g_{\mathrm{t\phi}} \Omega}{\sqrt{-g_{\mathrm{tt}} - 2 g_{\mathrm{t\phi}} \Omega - g_{\mathrm{\phi\phi}} \Omega^{2}}}, \label{dae} \\
		L &= \frac{g_{\mathrm{t\phi}} + g_{\mathrm{\phi\phi}} \Omega}{\sqrt{-g_{\mathrm{tt}} - 2 g_{\mathrm{t\phi}} \Omega - g_{\mathrm{\phi\phi}} \Omega^{2}}}. \label{dal}
	\end{align}
	
	The inner boundary of the thin accretion disk is determined by the innermost stable circular orbit (ISCO), which corresponds to the marginally stable solution of $ \partial_\mathrm{r}^2 V_{\mathrm{eff}} = 0 $. This condition results in the equation
	\begin{equation}
		E^{2} g_{\mathrm{\phi\phi,rr}} + 2 E L g_{\mathrm{t\phi,rr}} + L^{2} g_{\mathrm{tt,rr}} - \left(g_{\mathrm{t\phi}}^{2} - g_{\mathrm{tt}} g_{\mathrm{\phi\phi}}\right)_{\mathrm{,rr}} = 0.
		\label{efang}
	\end{equation}
	The root $ r = r_{\mathrm{isco}} $ of this equation defines the ISCO radius; for radii $ r < r_{\mathrm{isco}} $, the equatorial circular orbits are dynamically unstable and cannot support a steady accretion flow.

Throughout this section, the disk is described by the standard Novikov--Thorne model: the accreting matter moves on prograde circular equatorial geodesics, the inner edge of the disk coincides with the ISCO, where a zero-torque boundary condition is imposed, and the flux integral in Eq. (\ref{flux}) below is evaluated only for $r \geq r_{\mathrm{isco}}$. The extension of the emitting region down to the event horizon, with plunging dynamics inside the ISCO, is introduced only within the phenomenological, optically thin imaging model of Sec.~\ref{section4} and is not used in any of the Novikov--Thorne quantities computed here; the two treatments correspond to two distinct emission models and are kept strictly separate.
	
\begin{figure*}[htbp]
		\centering
		\begin{subfigure}{0.38\textwidth}
			%% [REVISION NOTE] Figs. Fa03/Fa05/Ta03/Ta05/vLa03/vLa05 must be regenerated for the PROGRADE branch
			%% (the previous versions corresponded to the retrograde branch of Eq. (3.6)); see response letter.
			\includegraphics[width=\linewidth]{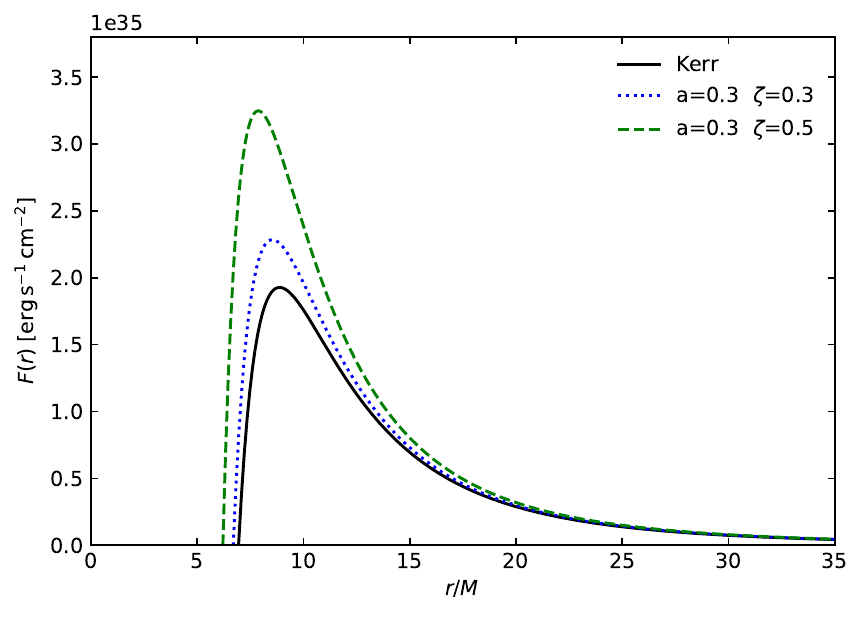}
		\end{subfigure}%
		\hspace{0.13\textwidth}% 
		\begin{subfigure}{0.38\textwidth}
			\includegraphics[width=\linewidth]{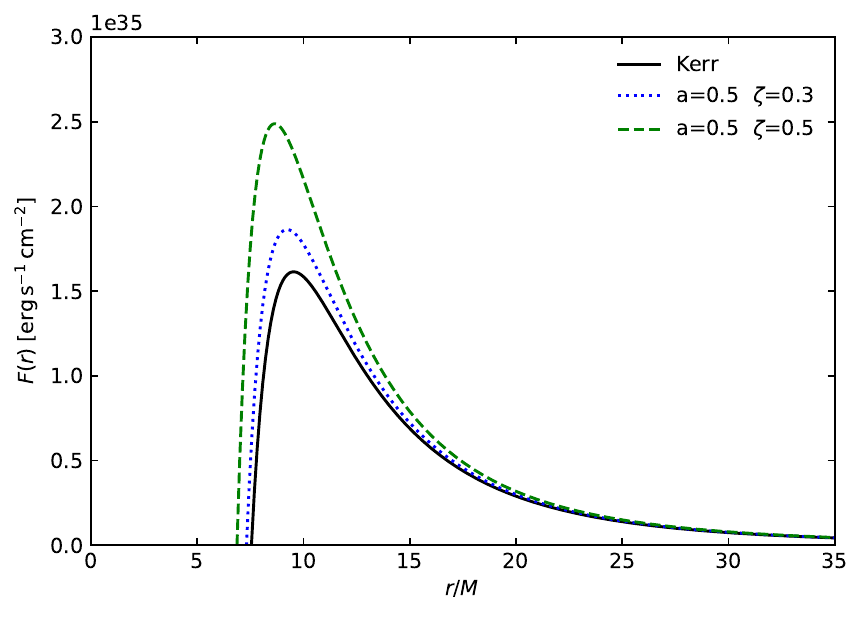}
		\end{subfigure}
		\caption{The energy flux $ F(r) $ from accretion disks around rotating Ay\'on-Beato--Garc\'{\i}a black holes for varying parameters $ a $ and $ \zeta $. Left panel: Fixed $ a = 0.3 $ ; Right panel: Fixed $ a = 0.5 $. The black lines represent the Kerr black hole case. All curves refer to prograde disks.}
		\label{Fr}
	\end{figure*}

	Accretion processes in thin disks surrounding rotating Ay\'on-Beato--Garc\'{\i}a black holes are next examined. The radiation flux emitted from the disk surface is determined by the conservation equations for the rest mass, the energy, and the angular momentum of the accreting fluid \cite{Page:1974he,NT}. Within the Novikov–Thorne model, the bolometric flux is expressed as
	\begin{equation}
		F(r) = -\frac{\dot{M}_{0}\,\Omega_{\mathrm{,r}}}{4\pi\sqrt{-g}\,(E - \Omega L)^{2}} \int_{r_{\mathrm{isco}}}^{r} (E - \Omega L)\,L_{\mathrm{,r}}\,\mathrm{d}r,
		\label{flux}
	\end{equation}
	where $\dot{M}_0$ refers to the mass accretion rate, and $g$ is defined as the determinant of the induced metric in the equatorial plane. As shown in Fig. \ref{Fr}, the energy flux $ F(r) $  exhibits a characteristic rise-and-fall behavior with radius, typical of Novikov–Thorne disks. Notably, the amplitude of the flux is enhanced with increasing charge parameter $\zeta$ when the spin $a$ is held constant: as shown in Table \ref{tab:isco}, increasing $\zeta$ moves the prograde ISCO inward and raises the radiative efficiency $\eta = 1 - E_{\mathrm{isco}}$, so that more gravitational binding energy is released by the disk. For the same reason, at fixed $\zeta$ the peak flux increases with the prograde spin $a$, in agreement with the well-known behavior of Novikov--Thorne disks around Kerr black holes. In the limit $\zeta \to 0$, our numerical results reproduce the analytic Kerr values of $r_{\mathrm{H}}$, $r_{\mathrm{isco}}$, and $\eta$, listed in the rows with $\zeta = 0$ of Table \ref{tab:isco}; this provides a direct benchmark validation of our implementation against the Kerr limit.
	
	\begin{table}[htbp]
	
	\centering
	\caption{Event horizon radius $r_{\mathrm{H}}$, prograde ISCO radius $r_{\mathrm{isco}}$, specific energy $E_{\mathrm{isco}}$, and radiative efficiency $\eta = 1 - E_{\mathrm{isco}}$ of the rotating ABG black hole for the parameter values used in Figs. \ref{Fr}--\ref{vLv} (in units of $M$). The rows with $\zeta = 0$ coincide with the analytic Kerr values, validating our numerical implementation.}
	\label{tab:isco}
	\begin{tabular}{cccccc}
	\hline\hline
	$a$ & $\zeta$ & $r_{\mathrm{H}}$ & $r_{\mathrm{isco}}$ & $E_{\mathrm{isco}}$ & $\eta$ \\
	\hline
	0.3 & 0 (Kerr) & 1.9539 & 4.9786 & 0.9306 & 0.0694 \\
	0.3 & 0.3      & 1.8256 & 4.6254 & 0.9259 & 0.0741 \\
	0.3 & 0.5      & 1.5221 & 3.8326 & 0.9128 & 0.0872 \\
	0.5 & 0 (Kerr) & 1.8660 & 4.2330 & 0.9179 & 0.0821 \\
	0.5 & 0.3      & 1.7168 & 3.7988 & 0.9094 & 0.0906 \\
	0.5 & 0.5      & 1.2584 & 2.5221 & 0.8723 & 0.1277 \\
	\hline\hline
	\end{tabular}
	\end{table}
	
\begin{figure*}[htbp]
		\centering
		\begin{subfigure}{0.38\textwidth}
			\includegraphics[width=\linewidth]{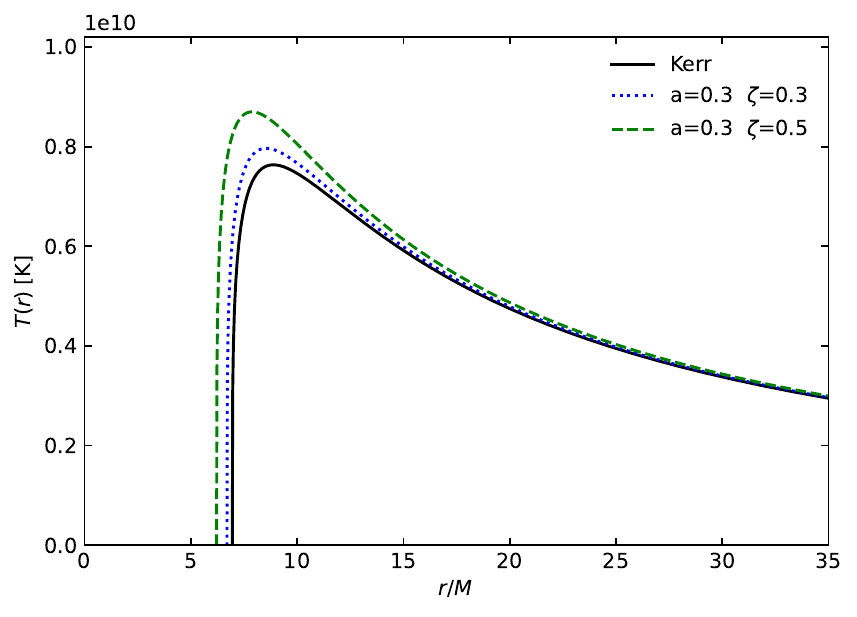}
		\end{subfigure}%
		\hspace{0.13\textwidth}% 
		\begin{subfigure}{0.38\textwidth}
			\includegraphics[width=\linewidth]{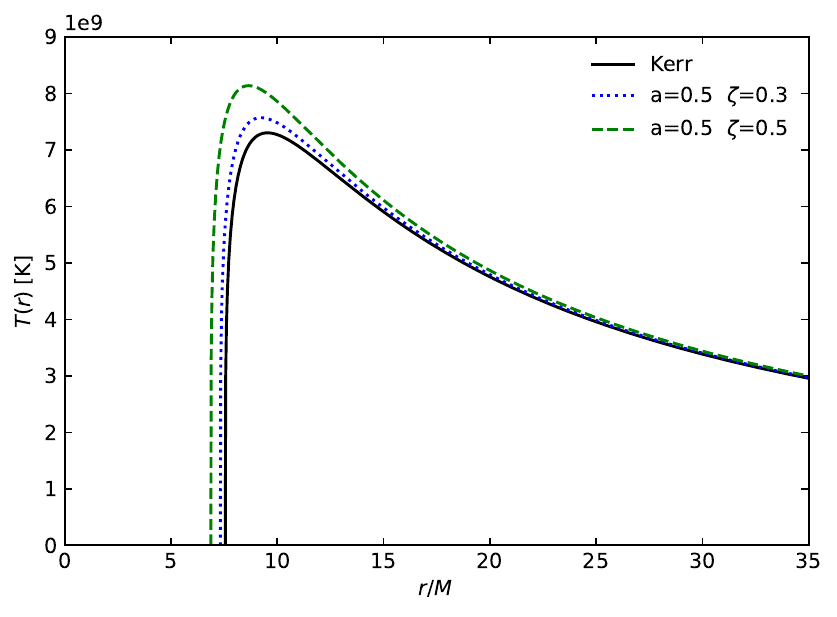}
		\end{subfigure}
		\caption{The radiation temperature $ T(r) $ from the accretion disk around a rotating Ay\'on-Beato--Garc\'{\i}a black hole for varying parameters $ a $ and $ \zeta $. Left panel: Fixed $ a = 0.3 $; Right panel: Fixed $ a = 0.5 $. The black lines represent the Kerr black hole case.}
		\label{Tr}
	\end{figure*}

	Using the Stefan–Boltzmann relation $ F(r) = \sigma_{\mathrm{SB}} T^{4}(r) $, we compute the radiation temperature $ T(r) $ of the thin disk. Fig. \ref{Tr} displays $ T(r) $ for different combinations of $ a $ and $ \zeta $. Like the energy flux, the temperature peaks at intermediate radii. Specifically, raising $\zeta$ at fixed $a$, or raising the prograde spin $a$ at fixed $\zeta$, enhances the peak temperature, again reflecting the inward shift of the ISCO and the corresponding increase of the radiative efficiency.
	
\begin{figure*}[htbp]
		\centering
		\begin{subfigure}{0.38\textwidth}
			\includegraphics[width=\linewidth]{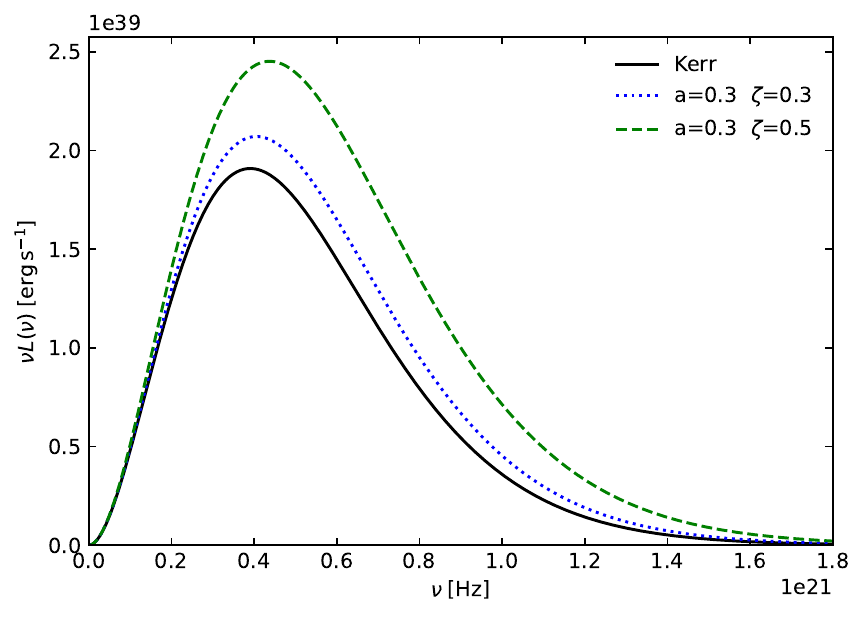}
		\end{subfigure}%
		\hspace{0.13\textwidth}% 
		\begin{subfigure}{0.38\textwidth}
			\includegraphics[width=\linewidth]{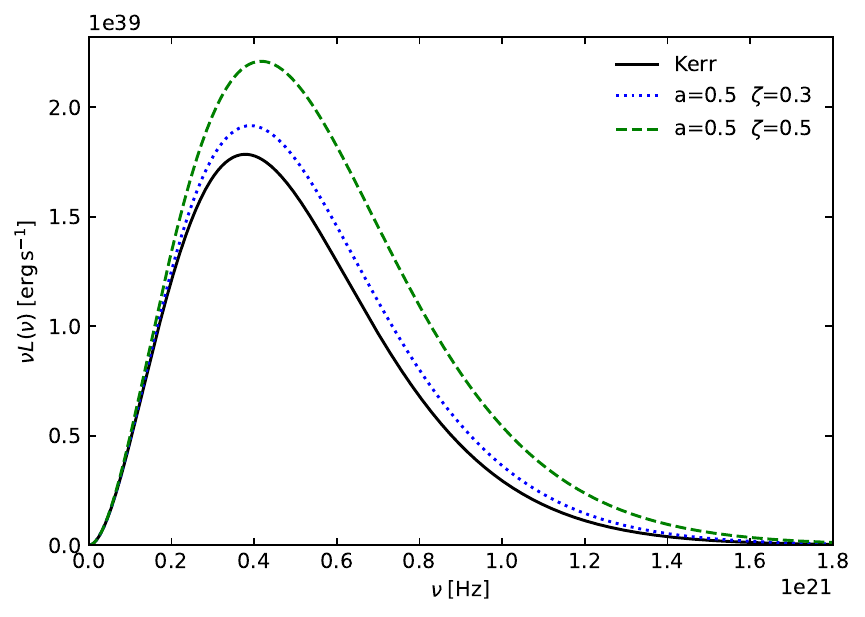}
		\end{subfigure}
		\caption{The spectral energy distribution $ \nu L(\nu) $ for thin accretion disks around rotating Ay\'on-Beato--Garc\'{\i}a black holes, shown for different values of the parameters $ a $ and $ \zeta $. Left panel: Fixed $ a = 0.3 $; Right panel: Fixed $ a = 0.5 $. The black lines represent the Kerr black hole case.}
		\label{vLv}
	\end{figure*}

	Following \cite{Torres:2002td}, the observed luminosity is
	\begin{equation}
		L(\nu) = \frac{8\pi h \cos\gamma}{c^{2}} \int_{r_{\mathrm{isco}}}^{\infty} \int_{0}^{2\pi} \frac{\nu_{\mathrm{e}}^{3} r}{e^{\mathrm{h}\nu_{\mathrm{e}}/(k_\mathrm{B} T)} - 1} \, \mathrm{d}\phi\, \mathrm{d}r,
		\label{17}
	\end{equation}
	with $ \nu_\mathrm{e} = \nu(1+z) $ and the redshift factor
	\begin{equation}
		1+z = \frac{1 + \Omega r \sin\gamma \sin\phi}{\sqrt{-g_{\mathrm{tt}} - 2g_{\mathrm{t\phi}}\Omega - g_{\mathrm{\phi\phi}}\Omega^{2}}}.
		\label{18}
	\end{equation}
	We set $ \gamma = 0 $ and neglect light bending \cite{Luminet:1979nyg}. The resulting spectral energy distribution, shown in Fig.~\ref{vLv}, mirrors the trends seen in $ F(r) $ and $ T(r) $: both a larger $\zeta$ and a larger prograde spin $a$ enhance the emission and shift the spectral peak to higher frequencies.
	
\section{Observational Images of Black Holes Illuminated by Thin Accretion Disks}\label{section4}	
The photons originating from the accretion disk and propagating to a distant observer are considered. Exploiting the spacetime symmetries in the time and azimuthal directions, the analysis is carried out in the local rest frame of a zero-angular-momentum observer (ZAMO) situated at $(t_\mathrm{o} = 0, r_\mathrm{o}, \theta_{\mathrm{obs}}, \phi_\mathrm{o} = 0)$. The corresponding orthonormal tetrad reads
	\begin{equation}
		\begin{split}
			&e_{(0)}=\left(\sqrt{\frac{g_{\mathrm{\phi\phi}}}{g_{\mathrm{t\phi}}^2-g_{\mathrm{tt}}g_{\mathrm{\phi\phi}}}}, 0, 0, -\frac{g_{\mathrm{t\phi}}}{g_{\mathrm{\phi\phi}}}\sqrt{\frac{g_{\mathrm{\phi\phi}}}{g_{\mathrm{t\phi}}^2-g_{\mathrm{tt}}g_{\mathrm{\phi\phi}}}}\right),\\
			&e_{(1)}=\left(0, -\frac{1}{\sqrt{g_{\mathrm{rr}}}}, 0, 0\right),\\
			&e_{(2)}=\left(0, 0, \frac{1}{\sqrt{g_{\mathrm{\theta\theta}}}}, 0\right),\\
			&e_{(3)}=\left(0, 0, 0, -\frac{1}{\sqrt{g_{\mathrm{\phi\phi}}}}\right),\label{zamo}
		\end{split}
	\end{equation}
A minus sign is incorporated into both $e_{(1)}$ and $e_{(2)}$ to facilitate the backward ray-tracing procedure, under which the photon paths are assumed reversible when emanating from the ZAMO. In the ZAMO's local frame, the four-momentum for photon is given by $p_{\mathrm{(\mu)}} = k_\mathrm{\nu} e^{\nu}_{\mathrm{(\mu)}}$. The celestial coordinates $(\Theta, \Psi)$, defined on the observer's image plane, with their relation to $p_{\mathrm{(\mu)}}$ are expressed as	
\begin{equation}
	\cos\Theta = \frac{p^{(1)}}{p^{(0)}}, \quad \tan\Psi = \frac{p^{(3)}}{p^{(2)}}.\label{4dl}
\end{equation}	
On the observer's screen—defined in the local rest frame of the ZAMO-we introduce a Cartesian coordinate system $(x, y)$ to map incoming light rays. In this setup,	
\begin{equation}
	x = -2\tan\frac{\Theta}{2}\sin\Psi, \quad y = -2\tan\frac{\Theta}{2}\cos\Psi.\label{xy}
\end{equation}	

During backward ray tracing, the photon trajectory is found to intersect the equatorial plane multiple times. The radii of these intersections are denoted by $r_\mathrm{n}(x, y)$, with $n = 1, 2, \dots, N_{\mathrm{max}}(x, y)$, where $N_{\mathrm{max}}(x, y)$ is the maximum number of crossings at screen coordinates $(x, y)$. The resulting discrete set $\{r_\mathrm{n}(x, y)\}$ constitutes the transfer function, which determines the shape of the $n$-th image of the disk—the primary image for $n=1$, the first lensed image for $n=2$, and so on. This function is known to depend on the observational angle $\theta_{\mathrm{obs}}$. A null geodesic linking the emission region (i.e., the disk) to the observer’s screen in the ZAMO frame is assumed. The variation of specific intensity along this ray is attributed to emission and absorption within the disk. Under the assumption of negligible refraction, the radiative transfer equation takes the form \cite{Lindquist:1966igj}:
\begin{equation}
\frac{\mathrm{d}}{\mathrm{d}\lambda}\left(\frac{I_\mathrm{\nu}}{\nu^3}\right) = \frac{J_\mathrm{\nu} - \kappa_\mathrm{\nu} I_\mathrm{\nu}}{\nu^2},\label{transeq}
\end{equation}
with $\lambda$ the affine parameter, and $I_\mathrm{\nu}$, $J_\mathrm{\nu}$, $\kappa_\mathrm{\nu}$ denoting the specific intensity, emissivity, and the absorption coefficient at frequency $\nu$, respectively. In vacuum, where both $J_\nu$ and $\kappa_\mathrm{\nu}$ vanish, the ratio $I_\mathrm{\nu} / \nu^3$ remains constant along the geodesic.

Let the accretion disk satisfy the following conditions: (i) stationarity, (ii) axisymmetry, (iii) $Z_2$ symmetry with respect to the equatorial plane, and (iv) geometric thinness such that $J_\mathrm{\nu}$ and $\kappa_\mathrm{\nu}$ are independent of the polar angle near $\theta = \pi/2$. Under these hypotheses, the solution to the general-relativistic radiative transfer Eq. (\ref{transeq}) along a backward-integrated null geodesic $\gamma$ is given by
\begin{equation}
	I_{\mathrm{\nu_o}}(x,y) = \sum_{n=1}^{N_{\mathrm{max}}(x,y)} \left( \frac{\nu_\mathrm{o}}{\nu_n} \right)^3 \frac{J_\mathrm{n}}{\tau_{\mathrm{n-1}}} \left[ \frac{1 - e^{-\kappa_\mathrm{n} f_\mathrm{n}}}{\kappa_\mathrm{n}} \right],\label{imuxy}
\end{equation}
where

\begin{itemize}
	\item $\nu_\mathrm{o} = -p_{(0)}|_{r=r_\mathrm{o}}$ is the photon frequency in the ZAMO frame at the screen;
	\item $\nu_\mathrm{n} = -k_\mathrm{\mu} u^\mathrm{\mu}|_{r=r_\mathrm{n}}$ is the frequency in the local rest frame $F_\mathrm{n}$ of the disk fluid at the $n$-th intersection radius $r_\mathrm{n}(x,y)$;
	\item $\tau_\mathrm{m}$ is the cumulative optical depth, defined by
	\[
	\tau_\mathrm{m} =
	\begin{cases}
		\exp\left( \sum_{\mathrm{k=1}}^{\mathrm{m}} \kappa_\mathrm{k} f_\mathrm{k} \right), & m \geq 1, \\
		1, & m = 0,
	\end{cases}
	\]
	\item $f_\mathrm{n} = \nu_\mathrm{n} \Delta\lambda_\mathrm{n}$ is a model-dependent weighting factor representing the effective path length through the emitting layer.
\end{itemize}	
The quantity $\Delta\lambda_\mathrm{n}$ in the fudge factor represents the variation of the affine parameter during the photon's passage through the disk medium associated with the $n$-th local frame $F_\mathrm{n}$. Under the assumption of an optically thin accretion flow—such that absorption may be neglected—the intensity formula (\ref{imuxy}) is reduced to
\begin{equation}
	I_{\nu_\mathrm{o}} = \sum_{n=1}^{N_{\mathrm{max}}} f_\mathrm{n} \, g_\mathrm{n}^3 \, J_\mathrm{n},\label{imu0}
\end{equation}
where the redshift factor is given by $g_\mathrm{n} = \nu_\mathrm{o} / \nu_\mathrm{n}$.

The intensity formula given in Eq. (\ref{imu0}) has been utilized in prior investigations of optically thin accretion disks. In this work, the emissivity is specified as 
\begin{equation}
J(r) = \exp\!\left( -\frac{1}{2} z^2 - 2z \right), \quad z = \log(r/r_\mathrm{H}),\label{jr}
\end{equation}
a functional form adopted from Ref. \cite{Chael:2021rjo} to fit 230 GHz observations of M87$^{*}$ and Sgr A$^{*}$. Although the original implementation included radius-dependent $f_\mathrm{n}$, these factors are here normalized to unity. The absolute calibration of $f_\mathrm{n}$ influences only the subdominant photon ring. With $J_\mathrm{n}$ and $f_\mathrm{n}$ defined as above, the synthetic image is constructed by integrating Eq. (\ref{imu0}) along backward-traced null geodesics.

For enhanced physical interpretability, we provide a more explicit expression for the redshift factor $g_\mathrm{n} = \nu_\mathrm{o} / \nu_\mathrm{n}$. The accretion flow is assumed to consist of electrically neutral matter moving along the timelike geodesics. For the radii $r_\mathrm{n} \geq r_{\mathrm{ISCO}}$, the motion is restricted to circular orbits with angular velocity $\Omega_\mathrm{n} = (u^\mathrm{\phi} / u^\mathrm{t})|_{r_\mathrm{n}}$. Under this assumption, $g_\mathrm{n}$ is expressed as
\begin{equation}
g_\mathrm{n} = \frac{e}{\chi (1 - \Omega_\mathrm{n} b)},\label{gn}
\end{equation}
where the parameters $b$, $e$, and $\chi$ are defined by
\begin{equation}
b = \frac{k_\mathrm{\phi}}{-k_\mathrm{t}}, \quad
e = \frac{p_{(0)}}{k_\mathrm{t}}, \quad
\chi = \left. \sqrt{ \frac{-1}{g_{\mathrm{tt}} + 2 g_{\mathrm{t\phi}} \Omega_\mathrm{n} + g_{\mathrm{\phi\phi}} \Omega_\mathrm{n}^2} } \right|_{r = r_\mathrm{n}}.\label{bezt}
\end{equation}
In this formulation, $b$ represents the photon impact parameter, $e$ is the ratio of the photon energy measured locally by the emitter to its conserved energy at infinity, and $\chi$ encodes the relativistic normalization of the emitter's four-velocity; we denote it by $\chi$ to avoid confusion with the impact parameter $\xi = L/E$ introduced in Sec.~\ref{section2}. For asymptotically flat spacetimes, $e \to 1$ in the limit $r_\mathrm{o} \to \infty$.

In the interior region of the ISCO ($r_\mathrm{n} < r_{\mathrm{ISCO}}$), the accretion flow is assumed to follow critical plunging orbits, for which the radial component of the four-velocity is given by $u_\mathrm{c}^\mathrm{r}$. Under this assumption, the redshift factor is expressed as
\begin{equation}
g_\mathrm{n} = -\frac{e}{\frac{u_\mathrm{c}^\mathrm{r} k_\mathrm{r}}{\mathcal{E}} + E_{\mathrm{ISCO}} \left( g^{\mathrm{tt}} - g^{\mathrm{t\phi}} b \right) + L_{\mathrm{ISCO}} \left( g^{\mathrm{\phi\phi}} b - g^{\mathrm{t\phi}} \right)},\label{gn2}
\end{equation}
with $u_\mathrm{c}^\mathrm{r}$, $g^{\mathrm{tt}}$, $g^{\mathrm{t\phi}}$, and $g^{\mathrm{\phi\phi}}$ all evaluated at $r = r_\mathrm{n}$.

\begin{figure*}[htbp]
	\centering
	\setlength{\tabcolsep}{1pt} 
	\begin{tabular}{cccc}
		\begin{minipage}[t]{0.24\textwidth}
			\centering
			\hbox{
				\begin{overpic}[width=1.0\textwidth]{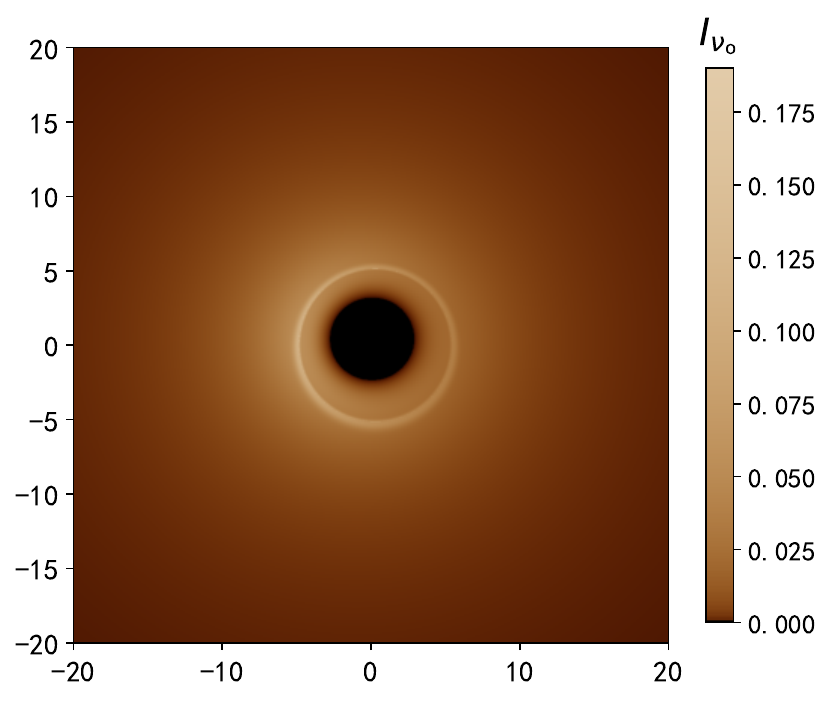}
					\put(15,85){\color{black} $a=0.5, \theta_{\mathrm{obs}}=17^{\circ}$} 
				\end{overpic}
			}
		\end{minipage}
		&
		\begin{minipage}[t]{0.24\textwidth}
			\centering
			\hbox{
				\begin{overpic}[width=1.0\textwidth]{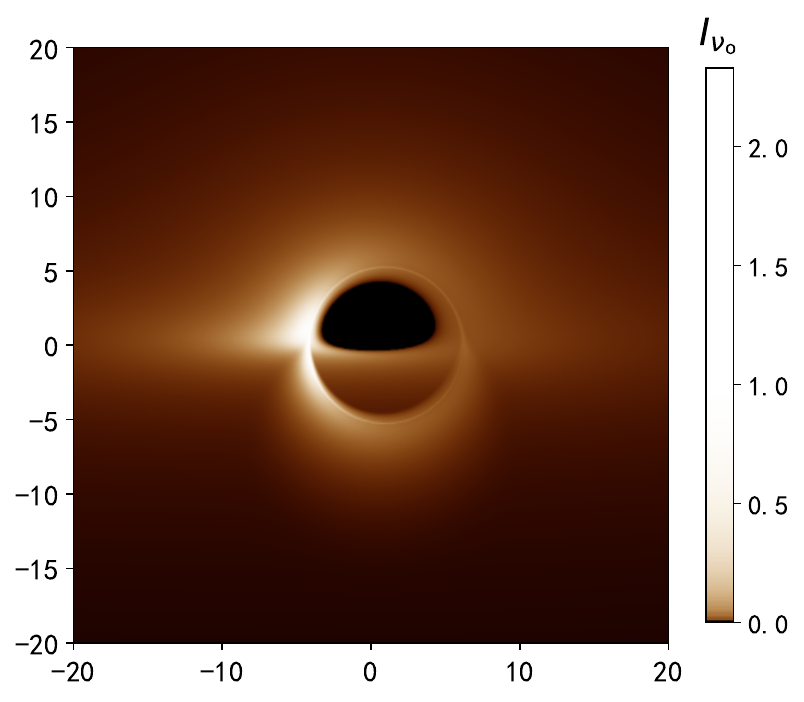}
					\put(15,85){\color{black} $a=0.5, \theta_{\mathrm{obs}}=80^{\circ}$}
				\end{overpic}
			}			
		\end{minipage}
		&
		\begin{minipage}[t]{0.24\textwidth}
			\centering
			\hbox{
				\begin{overpic}[width=1.0\textwidth]{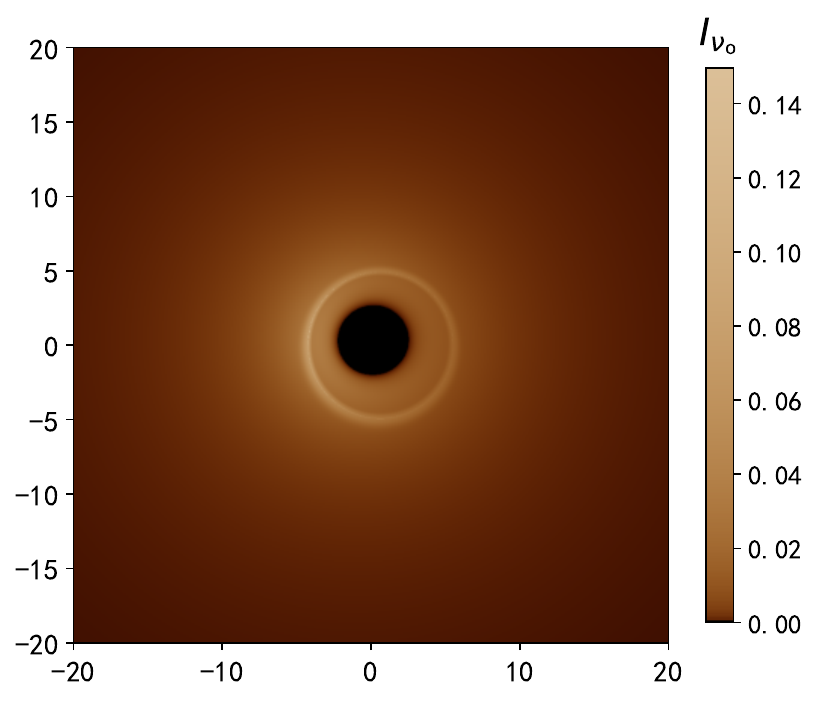}
					\put(15,85){\color{black} $a=0.95, \theta_{\mathrm{obs}}=17^{\circ}$}
				\end{overpic}
			}
		\end{minipage}
		&
		\begin{minipage}[t]{0.24\textwidth}
			\centering
			\hbox{
				\begin{overpic}[width=1.0\textwidth]{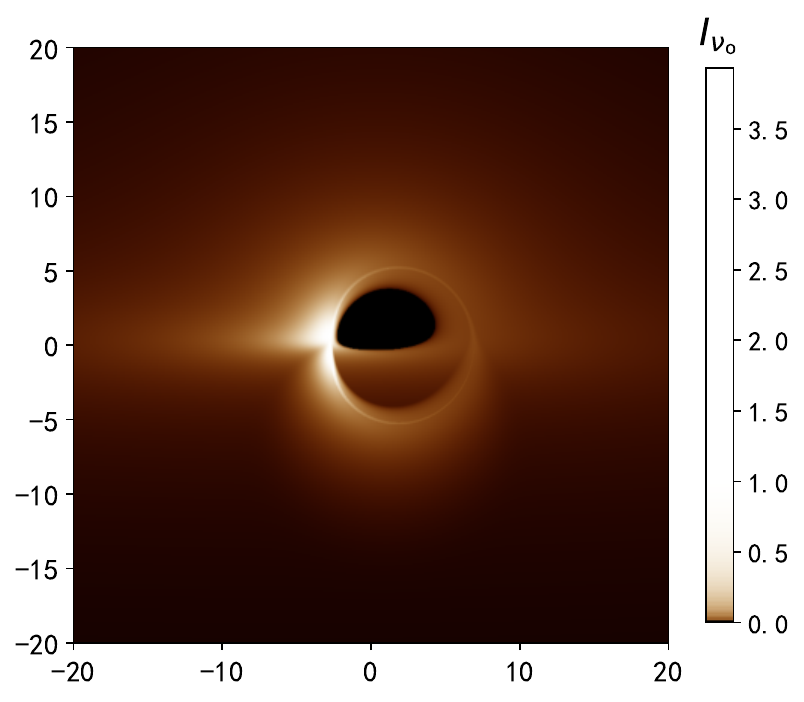}
					\put(15,85){\color{black} $a=0.95, \theta_{\mathrm{obs}}=80^{\circ}$}
				\end{overpic}
			}
		\end{minipage}
		\vspace{20pt} 
		\\ 
		\begin{minipage}[t]{0.24\textwidth}
			\centering
			\hbox{
				\begin{overpic}[width=1.0\textwidth]{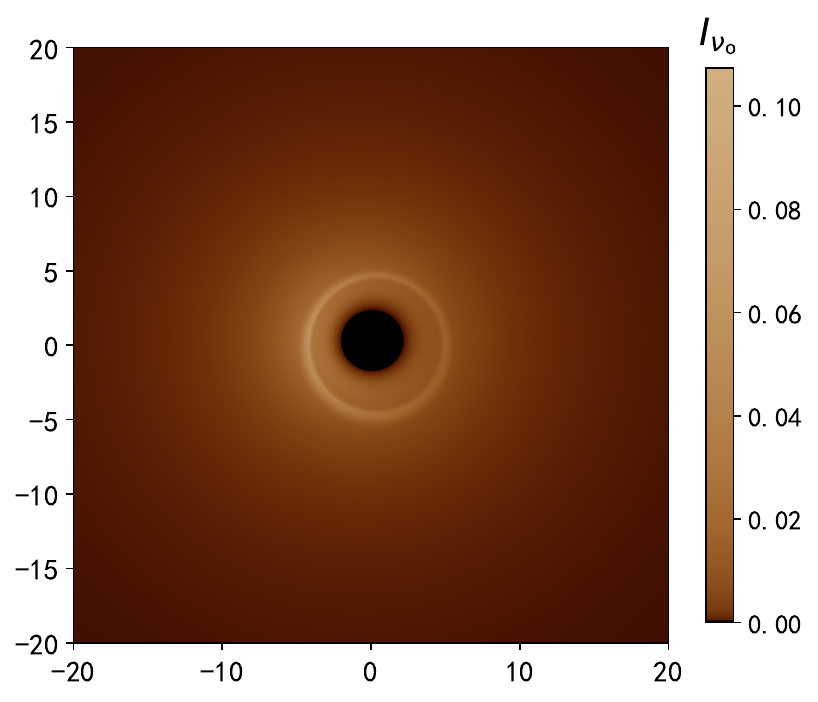}
					\put(3,88){\color{black} $a=0.5, \zeta=0.5, \theta_{\mathrm{obs}}=17^{\circ}$}
				\end{overpic}
			}
		\end{minipage}
		&
		\begin{minipage}[t]{0.24\textwidth}
			\centering
			\hbox{
				\begin{overpic}[width=1.0\textwidth]{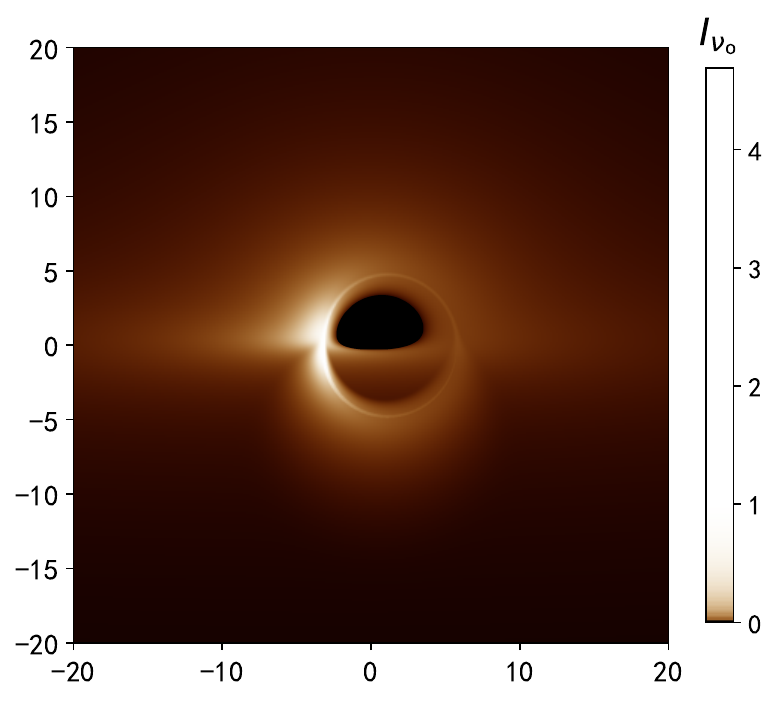}
					\put(3,88){\color{black} $a=0.5, \zeta=0.5, \theta_{\mathrm{obs}}=80^{\circ}$}
				\end{overpic}
			}
		\end{minipage}
		&
		\begin{minipage}[t]{0.24\textwidth}
			\centering
			\hbox{
				\begin{overpic}[width=1.0\textwidth]{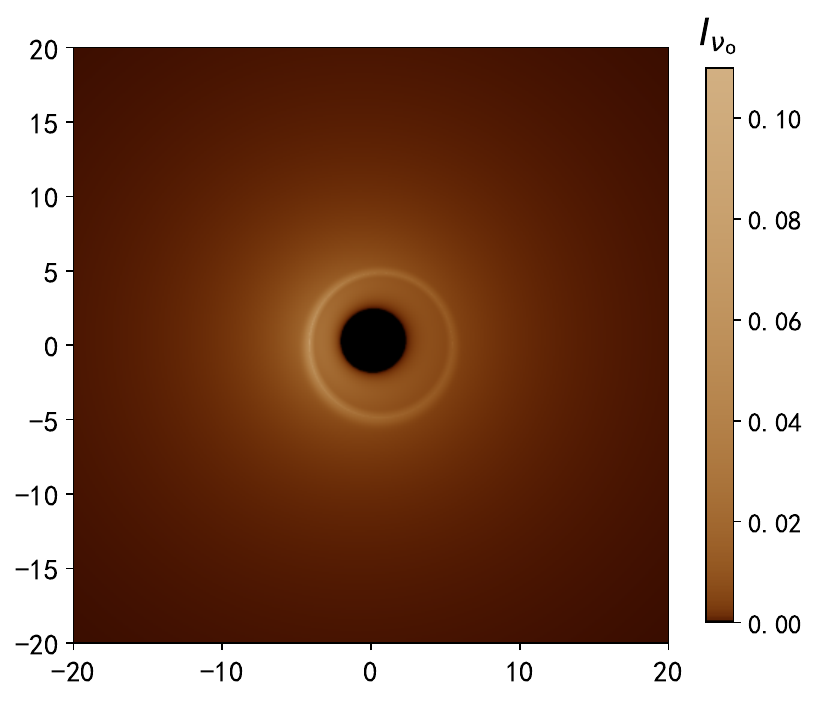}
					\put(1,88){\color{black} $a=0.95, \zeta=0.15, \theta_{\mathrm{obs}}=17^{\circ}$}
				\end{overpic}
			}
		\end{minipage}
		&
		\begin{minipage}[t]{0.24\textwidth}
			\centering
			\hbox{
				\begin{overpic}[width=1.0\textwidth]{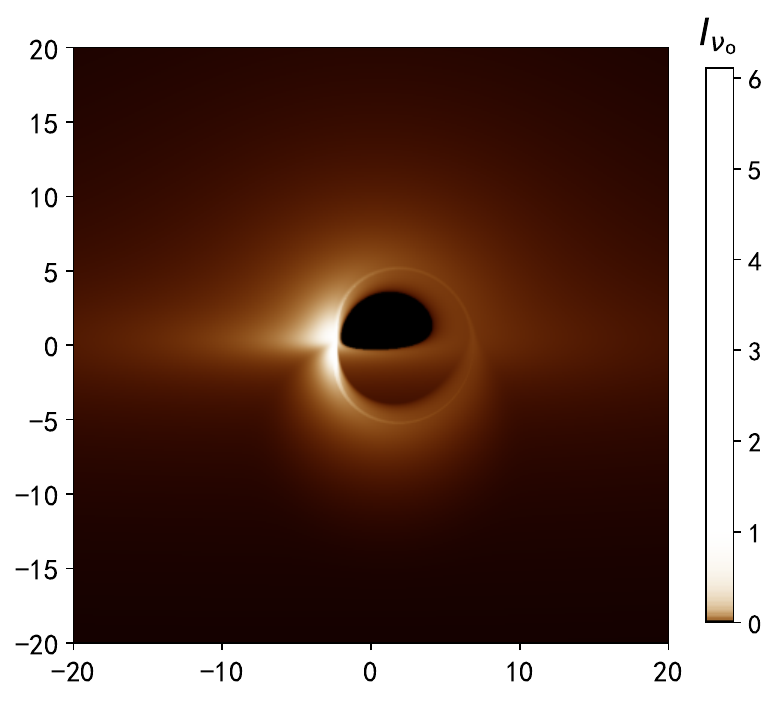}
					\put(1,88){\color{black} $a=0.95, \zeta=0.15, \theta_{\mathrm{obs}}=80^{\circ}$}
				\end{overpic}
			}
		\end{minipage}
	\end{tabular}
	\caption{Simulated images of a Kerr black hole (top) and a rotating Ay\'on-Beato--Garc\'{\i}a black hole (bottom), both surrounded by a prograde thin accretion disk.}
	\label{xijis}
\end{figure*}

\begin{figure*}[htbp]
	\centering
	\setlength{\tabcolsep}{1pt} 
	\begin{tabular}{cccc}
		\begin{minipage}[t]{0.24\textwidth}
			\centering
			\hbox{
				\begin{overpic}[width=1.0\textwidth]{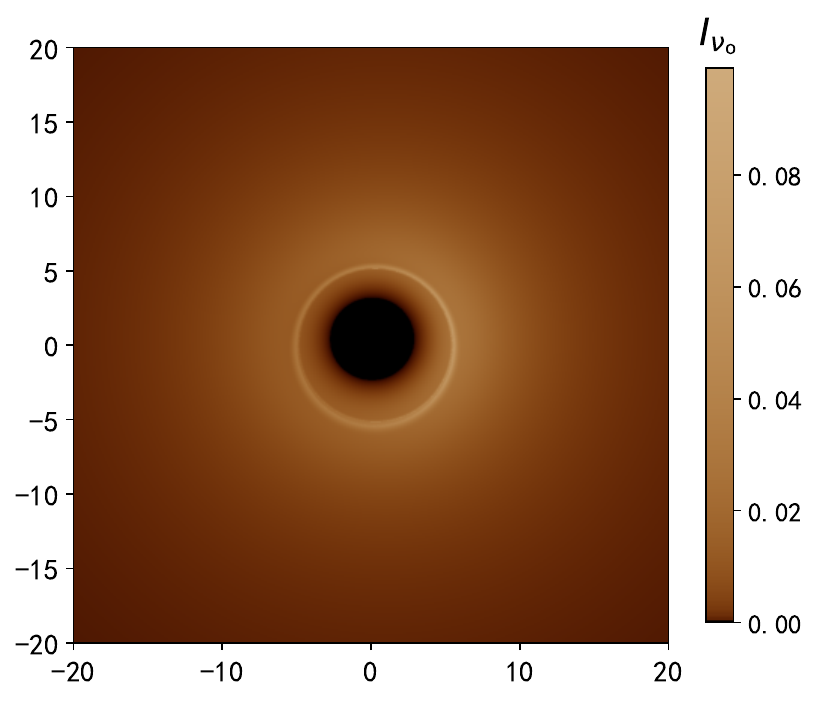}
					\put(15,85){\color{black} $a=0.5, \theta_{\mathrm{obs}}=17^{\circ}$} 
				\end{overpic}
			}
		\end{minipage}
		&
		\begin{minipage}[t]{0.24\textwidth}
			\centering
			\hbox{
				\begin{overpic}[width=1.0\textwidth]{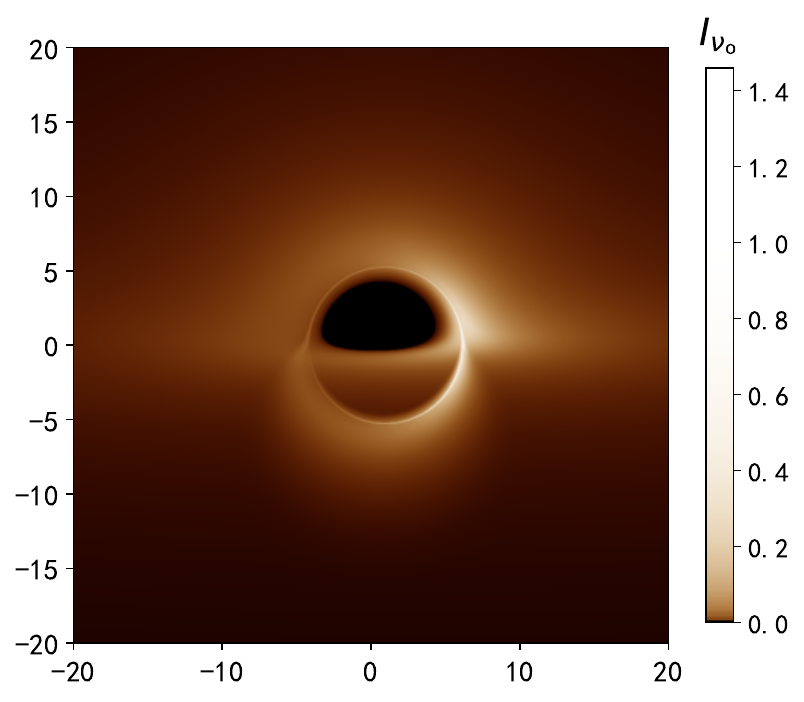}
					\put(15,85){\color{black} $a=0.5, \theta_{\mathrm{obs}}=80^{\circ}$}
				\end{overpic}
			}			
		\end{minipage}
		&
		\begin{minipage}[t]{0.24\textwidth}
			\centering
			\hbox{
				\begin{overpic}[width=1.0\textwidth]{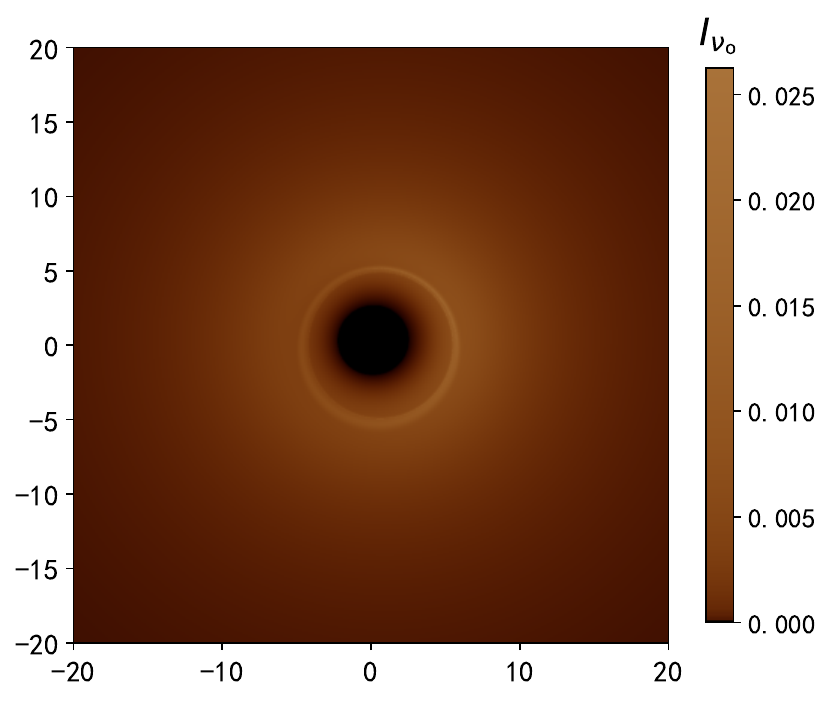}
					\put(15,85){\color{black} $a=0.95, \theta_{\mathrm{obs}}=17^{\circ}$}
				\end{overpic}
			}
		\end{minipage}
		&
		\begin{minipage}[t]{0.24\textwidth}
			\centering
			\hbox{
				\begin{overpic}[width=1.0\textwidth]{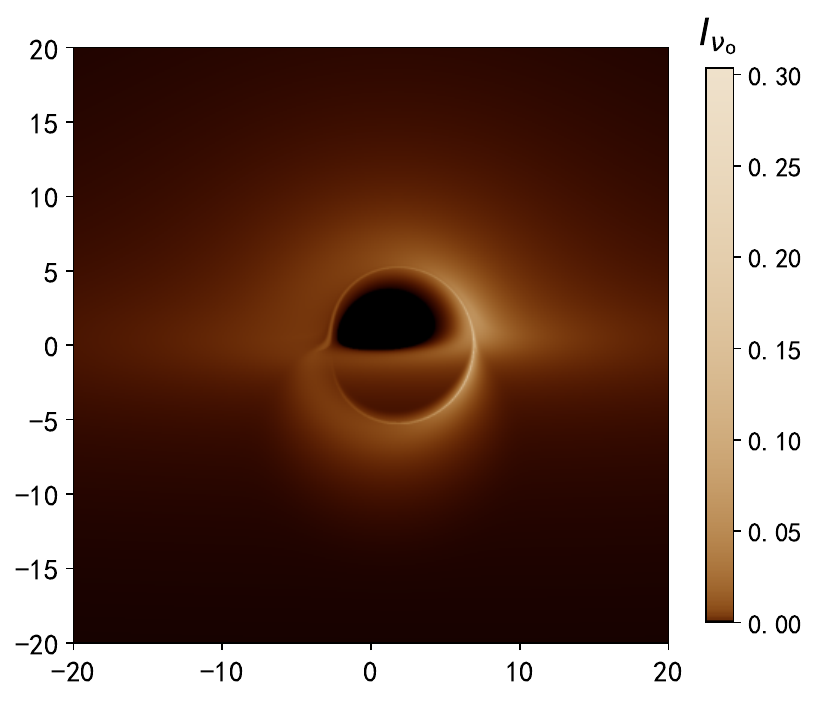}
					\put(15,85){\color{black} $a=0.95, \theta_{\mathrm{obs}}=80^{\circ}$}
				\end{overpic}
			}
		\end{minipage}
		\vspace{20pt} 
		\\ 
		\begin{minipage}[t]{0.24\textwidth}
			\centering
			\hbox{
				\begin{overpic}[width=1.0\textwidth]{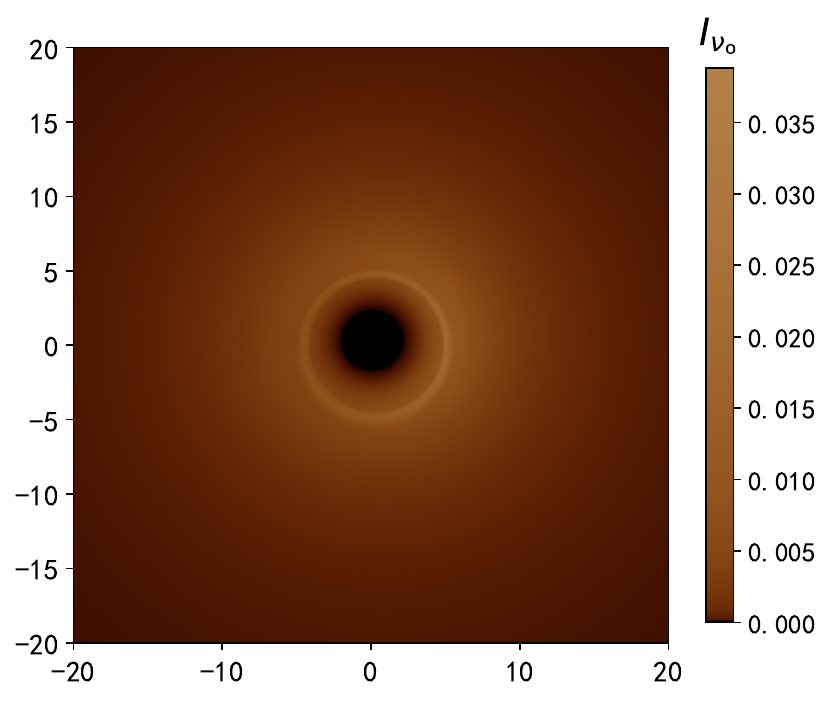}
					\put(3,85){\color{black} $a=0.5, \zeta=0.5, \theta_{\mathrm{obs}}=17^{\circ}$}
				\end{overpic}
			}
		\end{minipage}
		&
		\begin{minipage}[t]{0.24\textwidth}
			\centering
			\hbox{
				\begin{overpic}[width=1.0\textwidth]{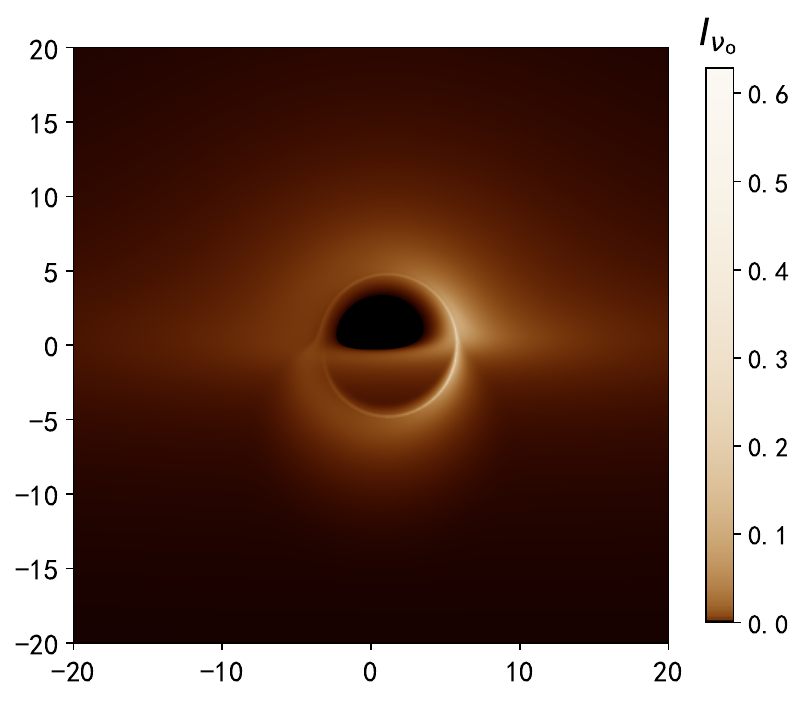}
					\put(3,85){\color{black} $a=0.5, \zeta=0.5, \theta_{\mathrm{obs}}=80^{\circ}$}
				\end{overpic}
			}
		\end{minipage}
		&
		\begin{minipage}[t]{0.24\textwidth}
			\centering
			\hbox{
				\begin{overpic}[width=1.0\textwidth]{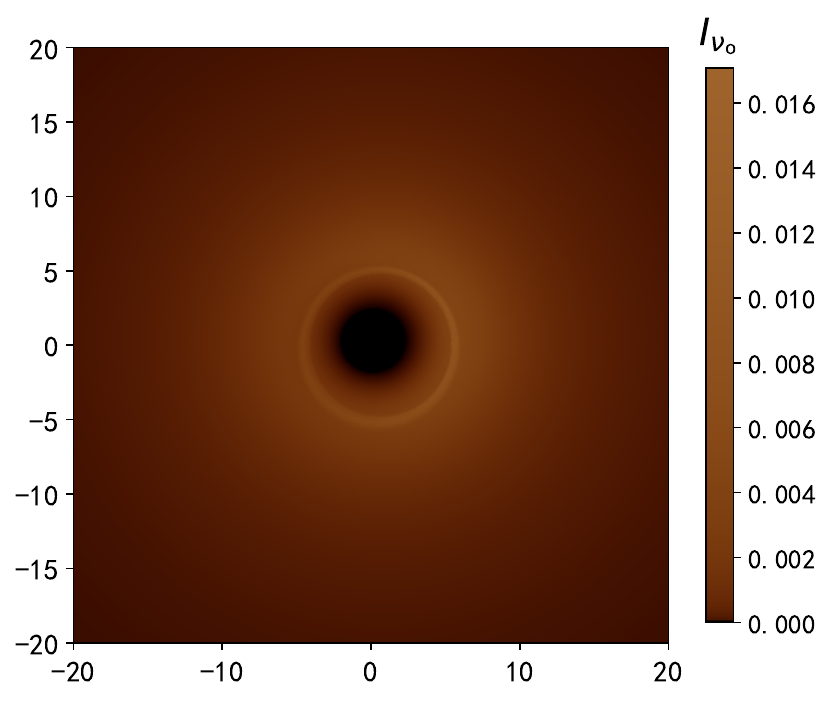}
					\put(1,85){\color{black} $a=0.95, \zeta=0.15, \theta_{\mathrm{obs}}=17^{\circ}$}
				\end{overpic}
			}
		\end{minipage}
		&
		\begin{minipage}[t]{0.24\textwidth}
			\centering
			\hbox{
				\begin{overpic}[width=1.0\textwidth]{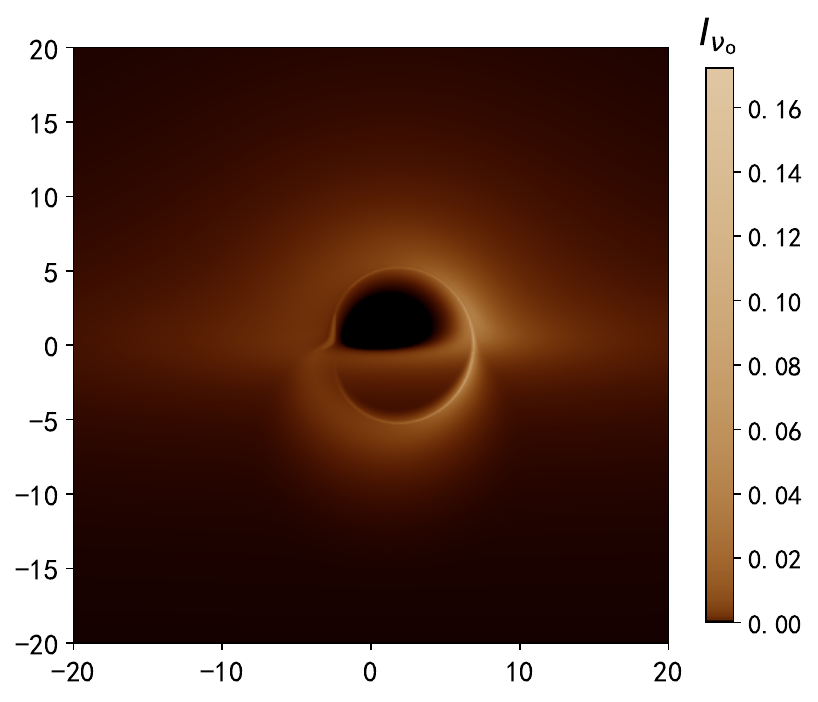}
					\put(1,85){\color{black} $a=0.95, \zeta=0.15, \theta_{\mathrm{obs}}=80^{\circ}$}
				\end{overpic}
			}
		\end{minipage}
	\end{tabular}
	\caption{Simulated images of a Kerr black hole (top) and a rotating Ay\'on-Beato--Garc\'{\i}a black hole (bottom), both illuminated by a retrograde thin accretion disk.}
	\label{xijin}
\end{figure*}

\begin{figure*}[htbp]
		\centering
		\setlength{\tabcolsep}{1pt} 
		\begin{tabular}{cccc}
			\begin{minipage}[t]{0.24\textwidth}
				\centering
				\hbox{
					\begin{overpic}[width=1.0\textwidth]{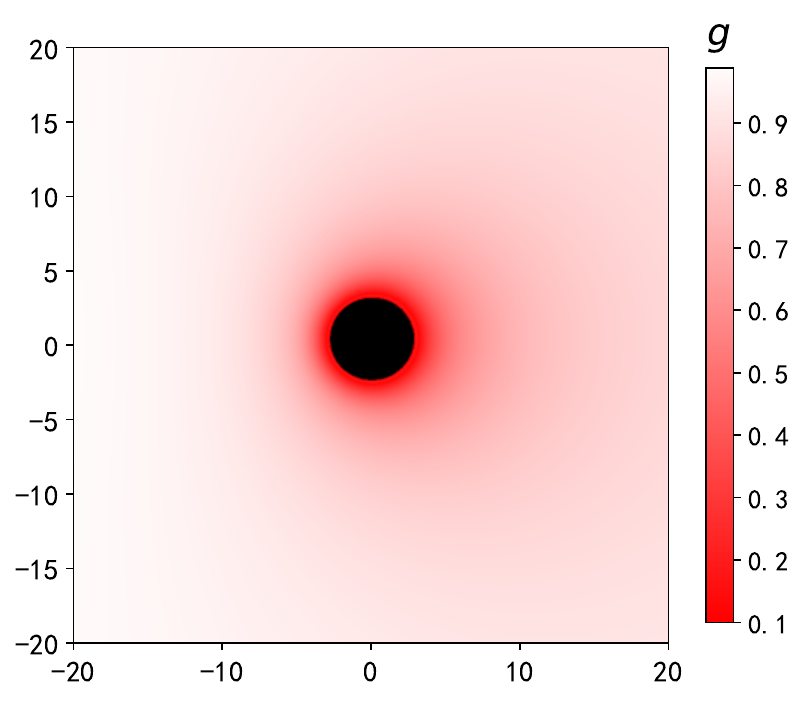}
						\put(15,85){\color{black} $a=0.5, \theta_{\mathrm{obs}}=17^{\circ}$} 
					\end{overpic}
				}
			\end{minipage}
			&
			\begin{minipage}[t]{0.24\textwidth}
				\centering
				\hbox{
					\begin{overpic}[width=1.0\textwidth]{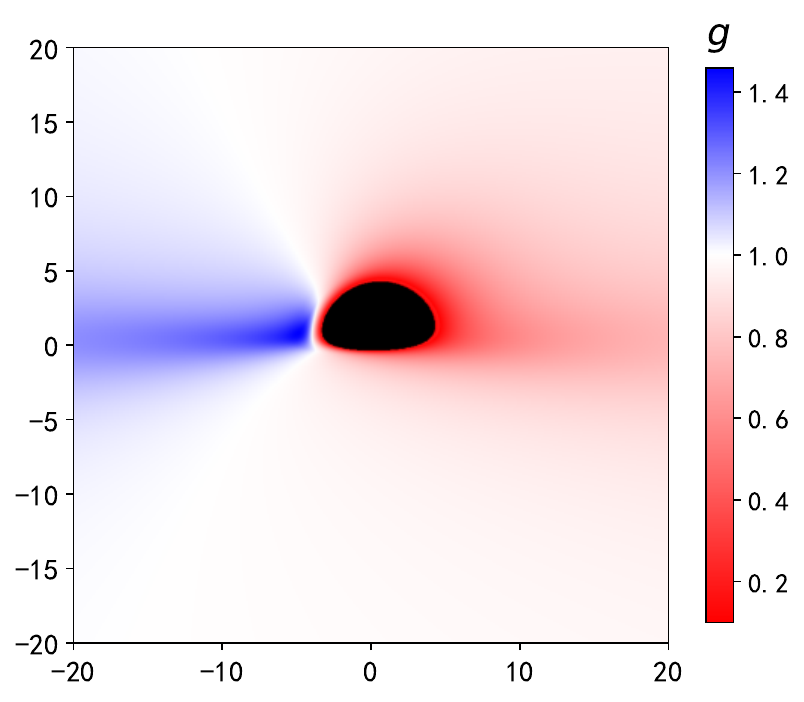}
						\put(15,85){\color{black} $a=0.5, \theta_{\mathrm{obs}}=80^{\circ}$}
					\end{overpic}
				}			
			\end{minipage}
			&
			\begin{minipage}[t]{0.24\textwidth}
				\centering
				\hbox{
					\begin{overpic}[width=1.0\textwidth]{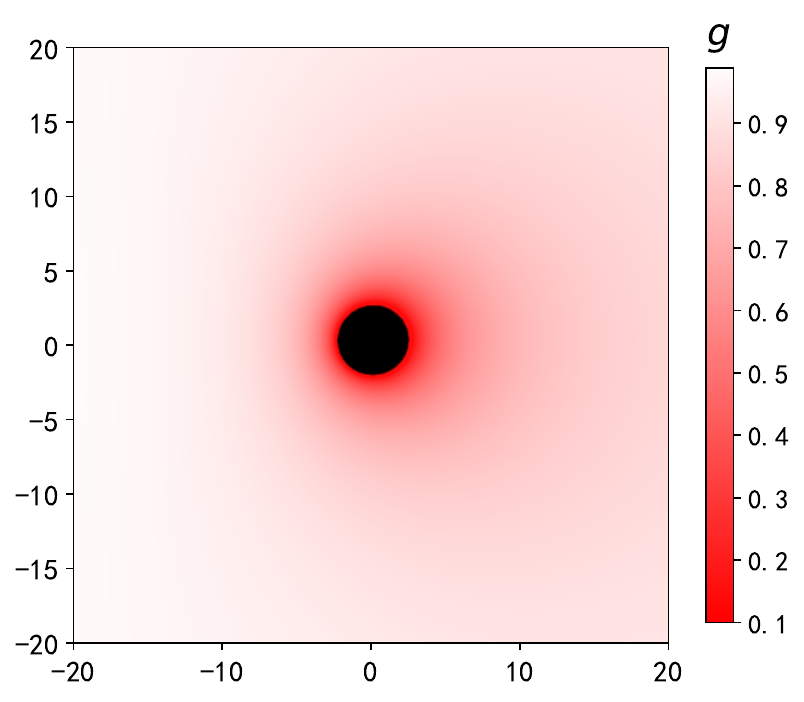}
						\put(15,85){\color{black} $a=0.95, \theta_{\mathrm{obs}}=17^{\circ}$}
					\end{overpic}
				}
			\end{minipage}
			&
			\begin{minipage}[t]{0.24\textwidth}
				\centering
				\hbox{
					\begin{overpic}[width=1.0\textwidth]{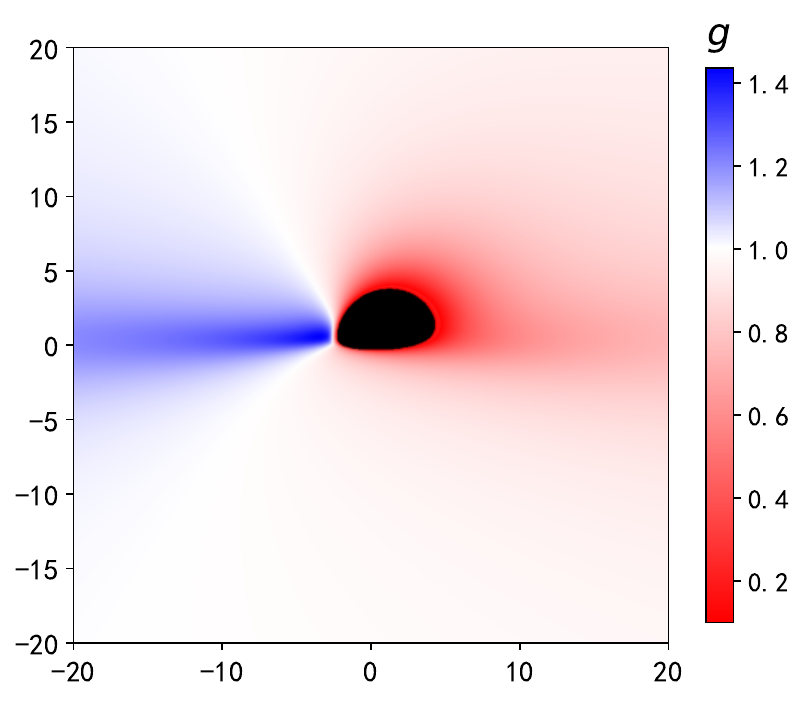}
						\put(15,85){\color{black} $a=0.95, \theta_{\mathrm{obs}}=80^{\circ}$}
					\end{overpic}
				}
			\end{minipage}
			\vspace{20pt} 
			\\ 
			\begin{minipage}[t]{0.24\textwidth}
				\centering
				\hbox{
					\begin{overpic}[width=1.0\textwidth]{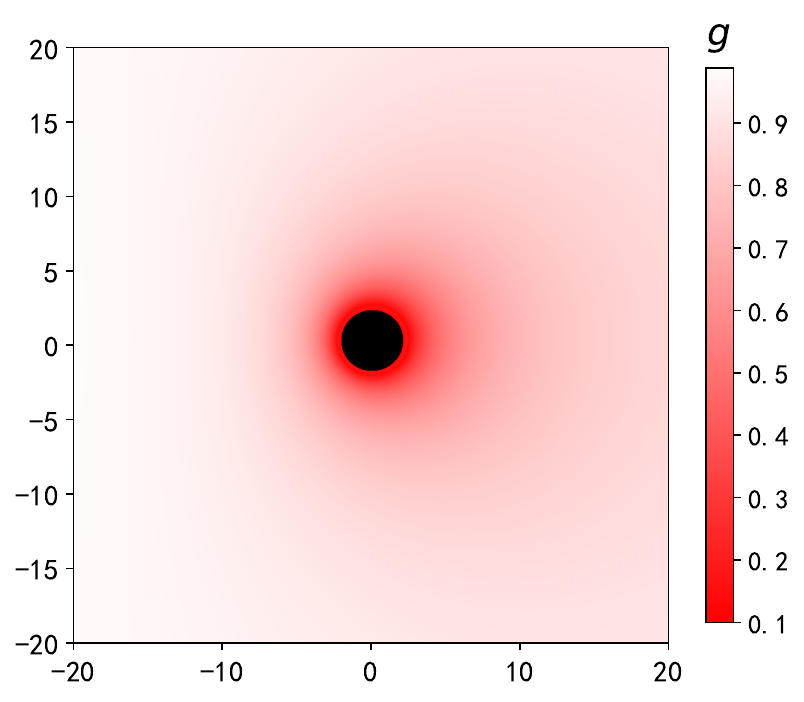}
						\put(3,85){\color{black} $a=0.5, \zeta=0.5, \theta_{\mathrm{obs}}=17^{\circ}$}
					\end{overpic}
				}
			\end{minipage}
			&
			\begin{minipage}[t]{0.24\textwidth}
				\centering
				\hbox{
					\begin{overpic}[width=1.0\textwidth]{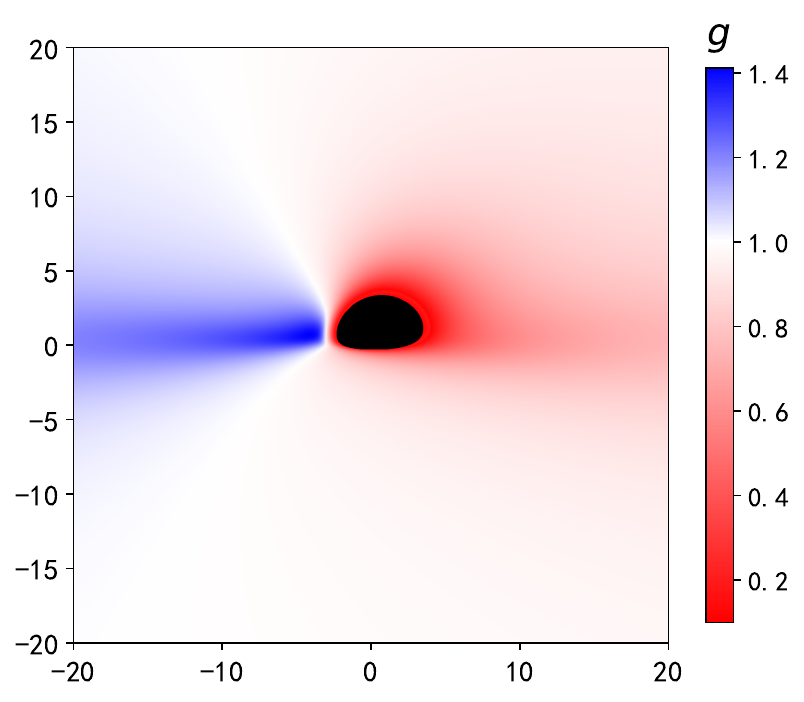}
						\put(3,85){\color{black} $a=0.5, \zeta=0.5, \theta_{\mathrm{obs}}=80^{\circ}$}
					\end{overpic}
				}
			\end{minipage}
			&
			\begin{minipage}[t]{0.24\textwidth}
				\centering
				\hbox{
					\begin{overpic}[width=1.0\textwidth]{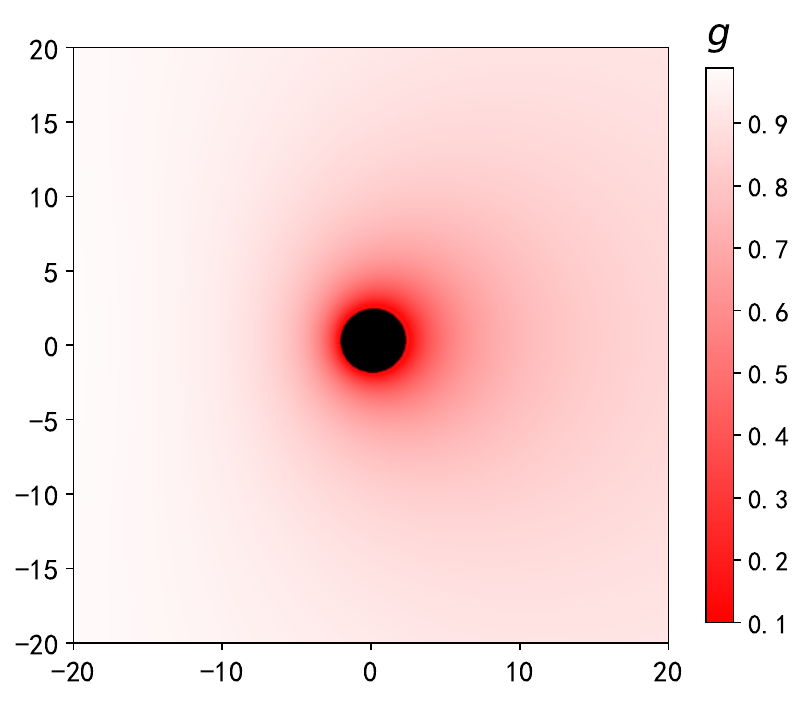}
						\put(1,85){\color{black} $a=0.95, \zeta=0.15, \theta_{\mathrm{obs}}=17^{\circ}$}
					\end{overpic}
				}
			\end{minipage}
			&
			\begin{minipage}[t]{0.24\textwidth}
				\centering
				\hbox{
					\begin{overpic}[width=1.0\textwidth]{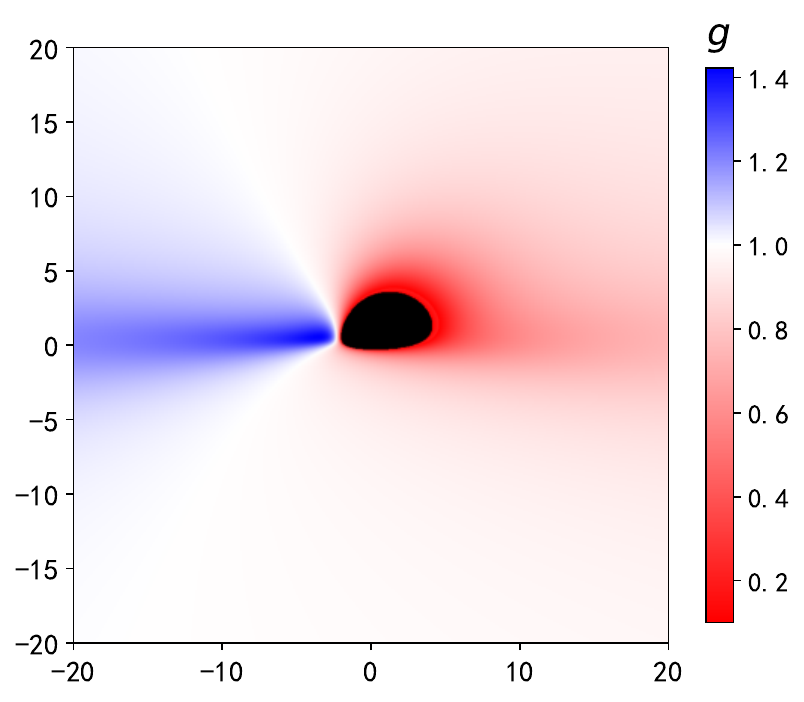}
						\put(1,85){\color{black} $a=0.95, \zeta=0.15, \theta_{\mathrm{obs}}=80^{\circ}$}
					\end{overpic}
				}
			\end{minipage}
		\end{tabular}
		\caption{The redshift distribution of the first-order (direct) image from prograde thin accretion disks, illustrating the dependence on spacetime parameters and viewing angle. The upper panel corresponds to a Kerr black hole, while the lower panel shows the rotating Ay\'on-Beato--Garc\'{\i}a solution.}
		\label{hongyis0}
	\end{figure*}

\begin{figure*}[htbp]
		\centering
		\setlength{\tabcolsep}{1pt} 
		\begin{tabular}{cccc}
			\begin{minipage}[t]{0.24\textwidth}
				\centering
				\hbox{
					\begin{overpic}[width=1.0\textwidth]{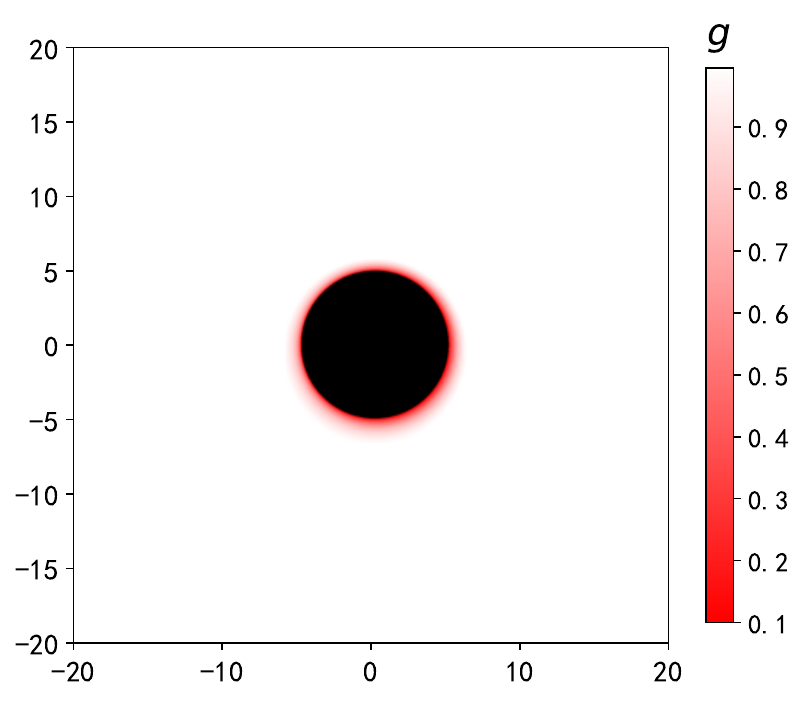}
						\put(15,85){\color{black} $a=0.5, \theta_{\mathrm{obs}}=17^{\circ}$} 
					\end{overpic}
				}
			\end{minipage}
			&
			\begin{minipage}[t]{0.24\textwidth}
				\centering
				\hbox{
					\begin{overpic}[width=1.0\textwidth]{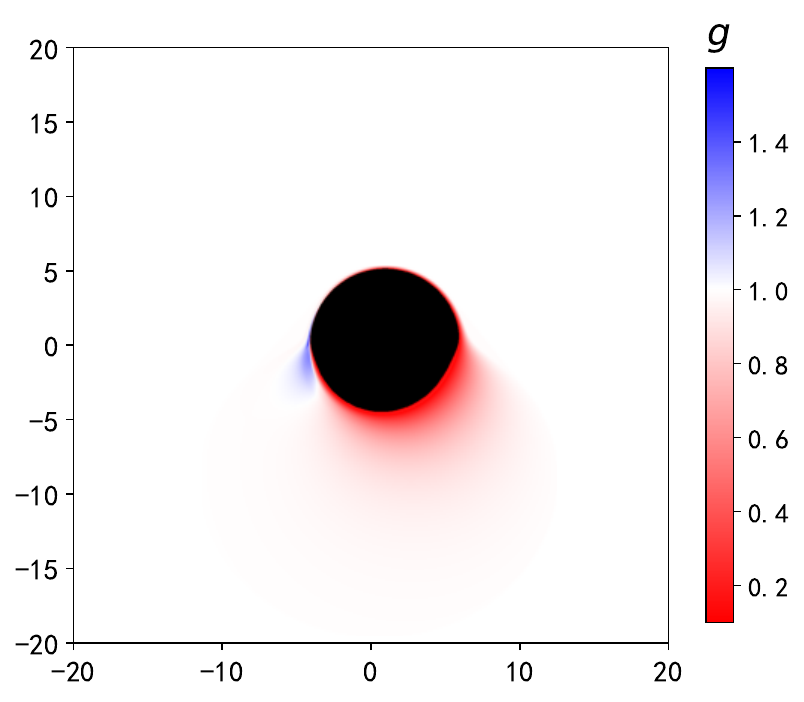}
						\put(15,85){\color{black} $a=0.5, \theta_{\mathrm{obs}}=80^{\circ}$}
					\end{overpic}
				}			
			\end{minipage}
			&
			\begin{minipage}[t]{0.24\textwidth}
				\centering
				\hbox{
					\begin{overpic}[width=1.0\textwidth]{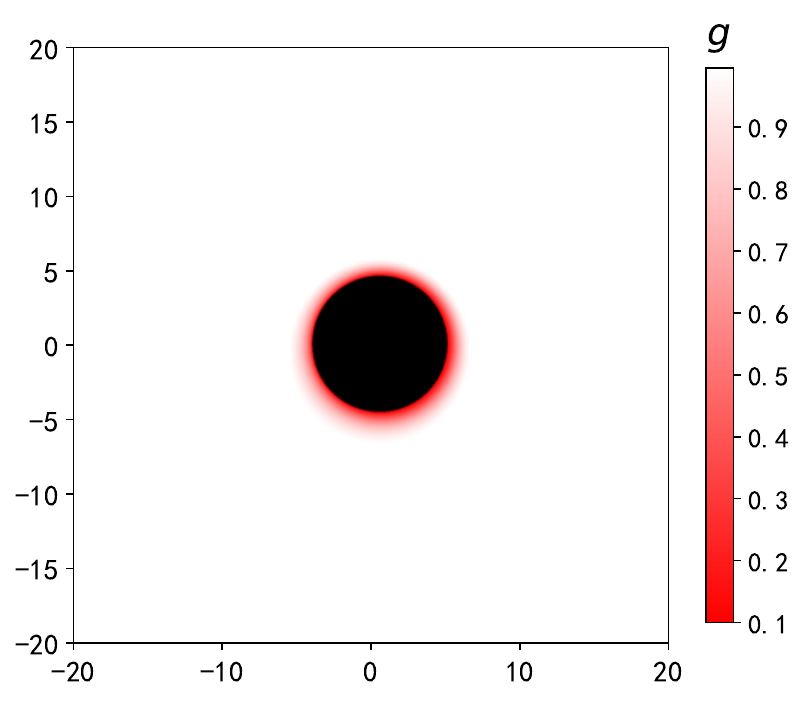}
						\put(15,85){\color{black} $a=0.95, \theta_{\mathrm{obs}}=17^{\circ}$}
					\end{overpic}
				}
			\end{minipage}
			&
			\begin{minipage}[t]{0.24\textwidth}
				\centering
				\hbox{
					\begin{overpic}[width=1.0\textwidth]{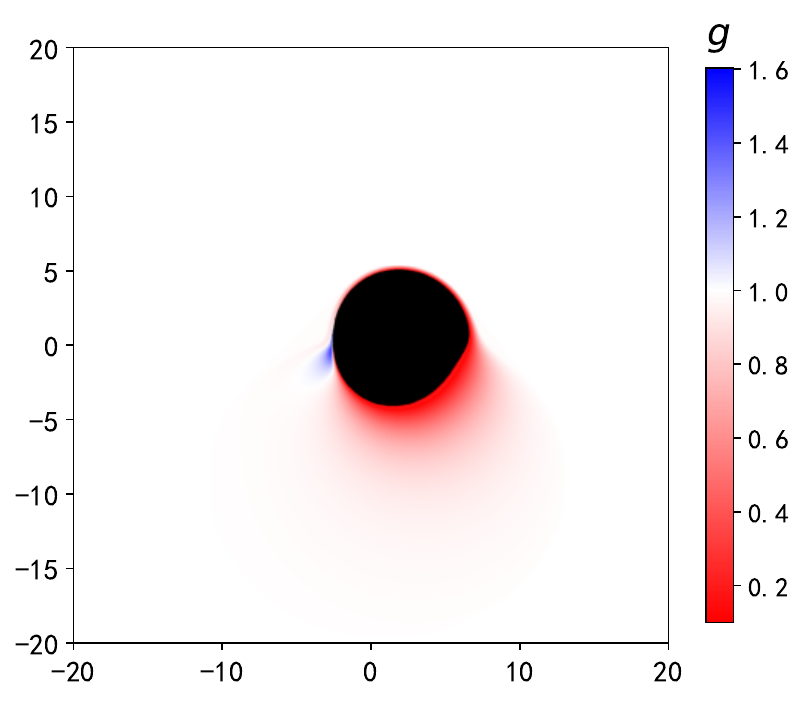}
						\put(15,85){\color{black} $a=0.95, \theta_{\mathrm{obs}}=80^{\circ}$}
					\end{overpic}
				}
			\end{minipage}
			\vspace{20pt} 
			\\ 
			\begin{minipage}[t]{0.24\textwidth}
				\centering
				\hbox{
					\begin{overpic}[width=1.0\textwidth]{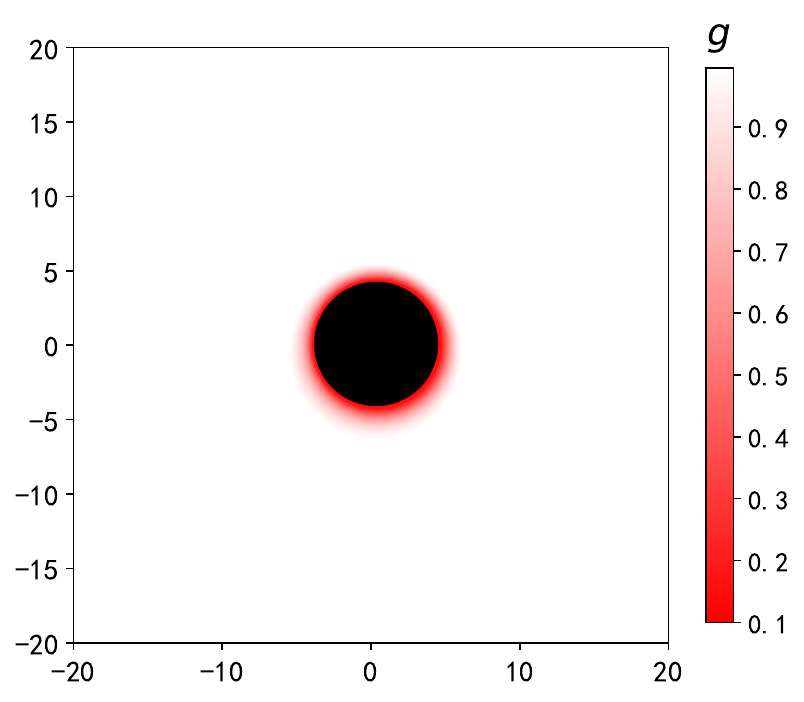}
						\put(3,85){\color{black} $a=0.5, \zeta=0.5, \theta_{\mathrm{obs}}=17^{\circ}$}
					\end{overpic}
				}
			\end{minipage}
			&
			\begin{minipage}[t]{0.24\textwidth}
				\centering
				\hbox{
					\begin{overpic}[width=1.0\textwidth]{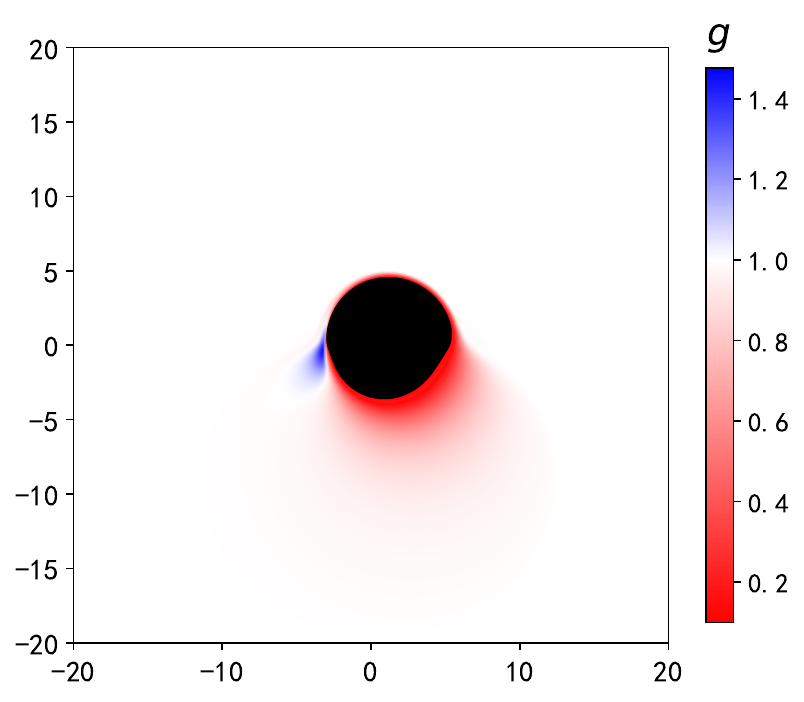}
						\put(3,85){\color{black} $a=0.5, \zeta=0.5, \theta_{\mathrm{obs}}=80^{\circ}$}
					\end{overpic}
				}
			\end{minipage}
			&
			\begin{minipage}[t]{0.24\textwidth}
				\centering
				\hbox{
					\begin{overpic}[width=1.0\textwidth]{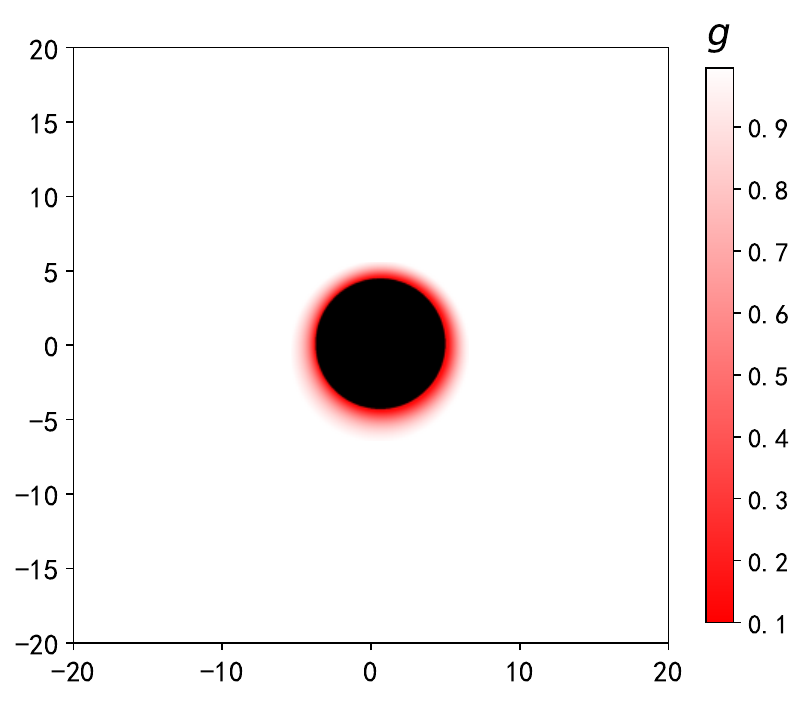}
						\put(1,85){\color{black} $a=0.95, \zeta=0.15, \theta_{\mathrm{obs}}=17^{\circ}$}
					\end{overpic}
				}
			\end{minipage}
			&
			\begin{minipage}[t]{0.24\textwidth}
				\centering
				\hbox{
					\begin{overpic}[width=1.0\textwidth]{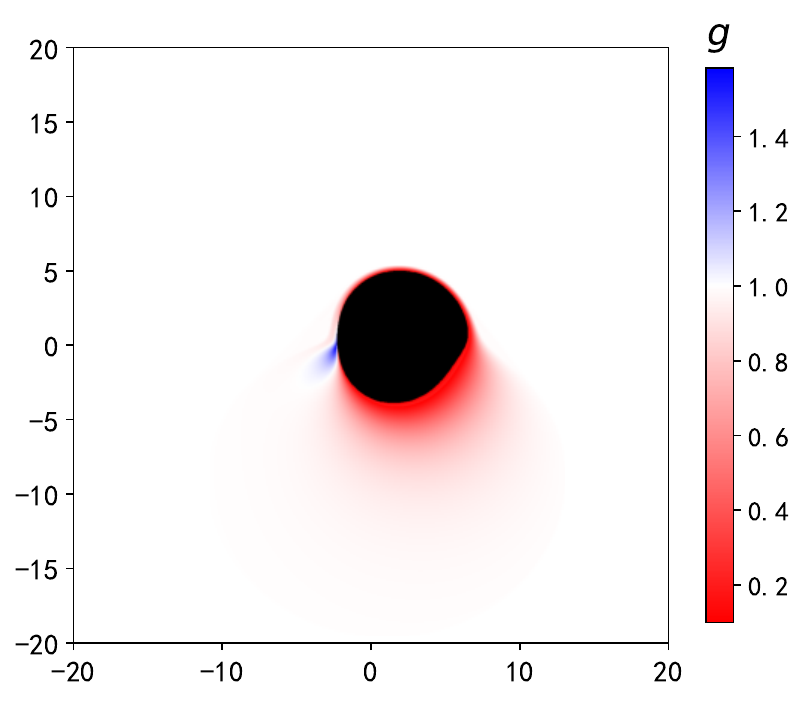}
						\put(1,85){\color{black} $a=0.95, \zeta=0.15, \theta_{\mathrm{obs}}=80^{\circ}$}
					\end{overpic}
				}
			\end{minipage}
		\end{tabular}
		\caption{The redshift distribution of the second-order (lensed) image from prograde thin accretion disks, illustrating the dependence on spacetime parameters and viewing angle. The upper panel corresponds to a Kerr black hole, while the lower panel shows the rotating Ay\'on-Beato--Garc\'{\i}a solution.}
		\label{hongyis1}
	\end{figure*}

\begin{figure*}[htbp]
		\centering
		\setlength{\tabcolsep}{1pt} 
		\begin{tabular}{cccc}
			\begin{minipage}[t]{0.24\textwidth}
				\centering
				\hbox{
					\begin{overpic}[width=1.0\textwidth]{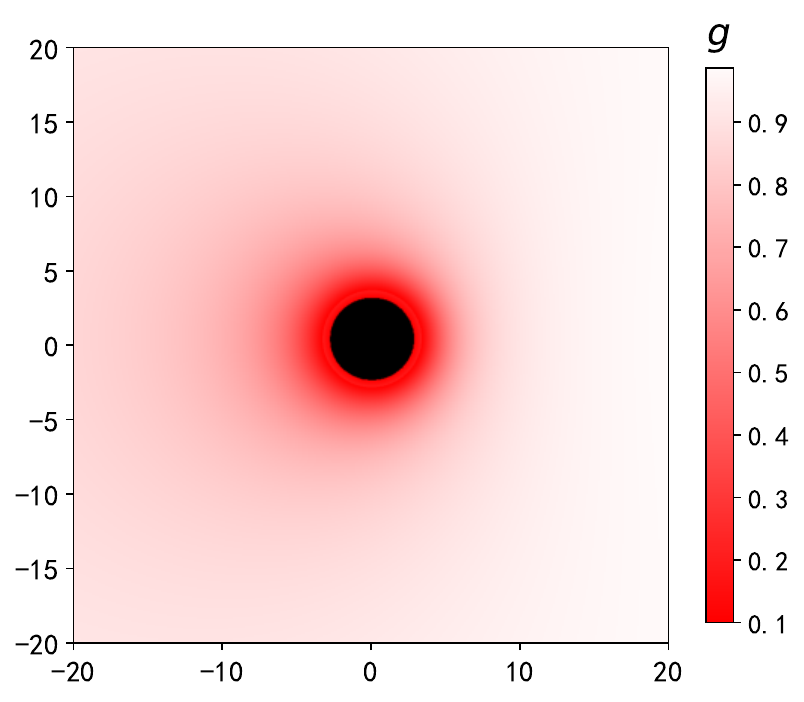}
						\put(15,85){\color{black} $a=0.5, \theta_{\mathrm{obs}}=17^{\circ}$} 
					\end{overpic}
				}
			\end{minipage}
			&
			\begin{minipage}[t]{0.24\textwidth}
				\centering
				\hbox{
					\begin{overpic}[width=1.0\textwidth]{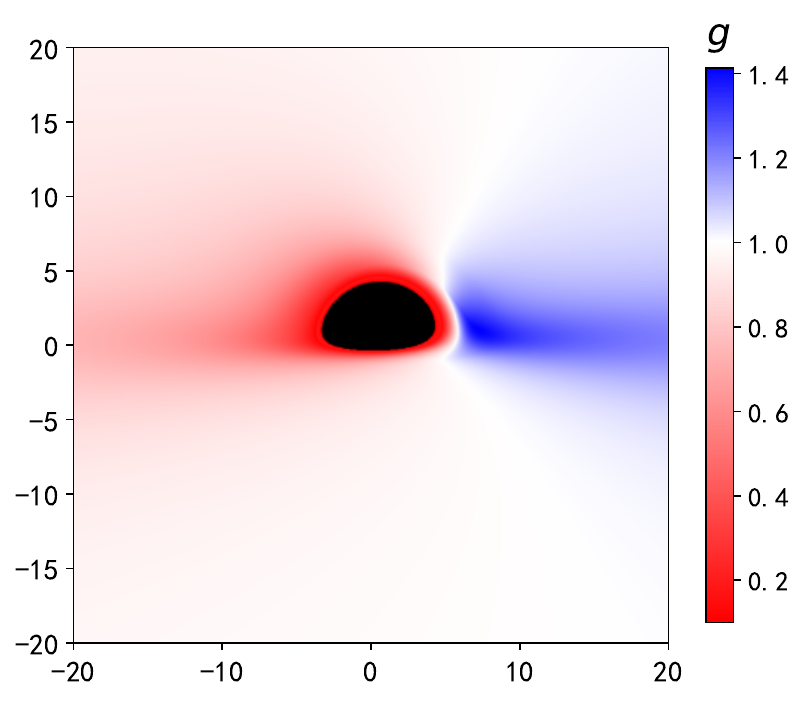}
						\put(15,85){\color{black} $a=0.5, \theta_{\mathrm{obs}}=80^{\circ}$}
					\end{overpic}
				}			
			\end{minipage}
			&
			\begin{minipage}[t]{0.24\textwidth}
				\centering
				\hbox{
					\begin{overpic}[width=1.0\textwidth]{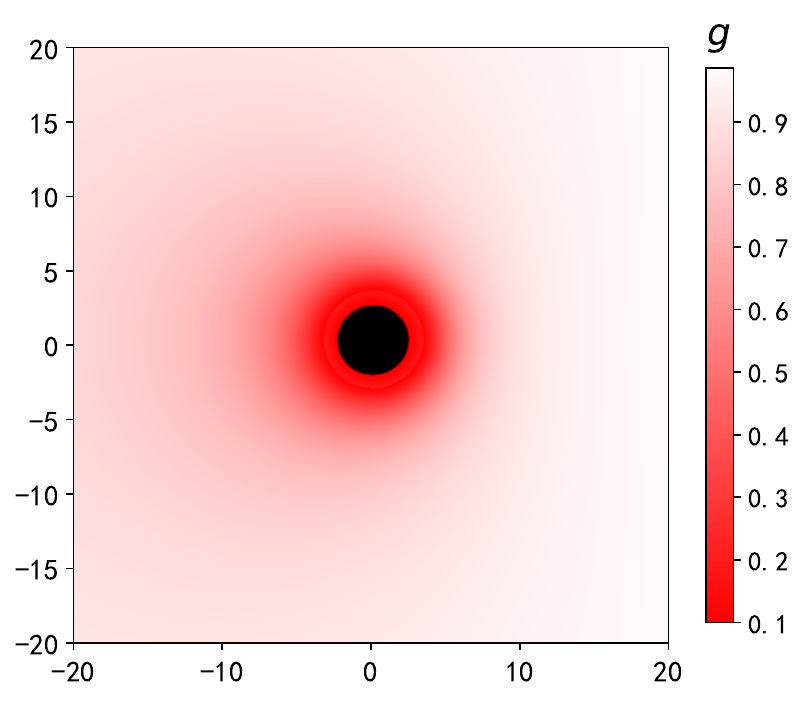}
						\put(15,85){\color{black} $a=0.95, \theta_{\mathrm{obs}}=17^{\circ}$}
					\end{overpic}
				}
			\end{minipage}
			&
			\begin{minipage}[t]{0.24\textwidth}
				\centering
				\hbox{
					\begin{overpic}[width=1.0\textwidth]{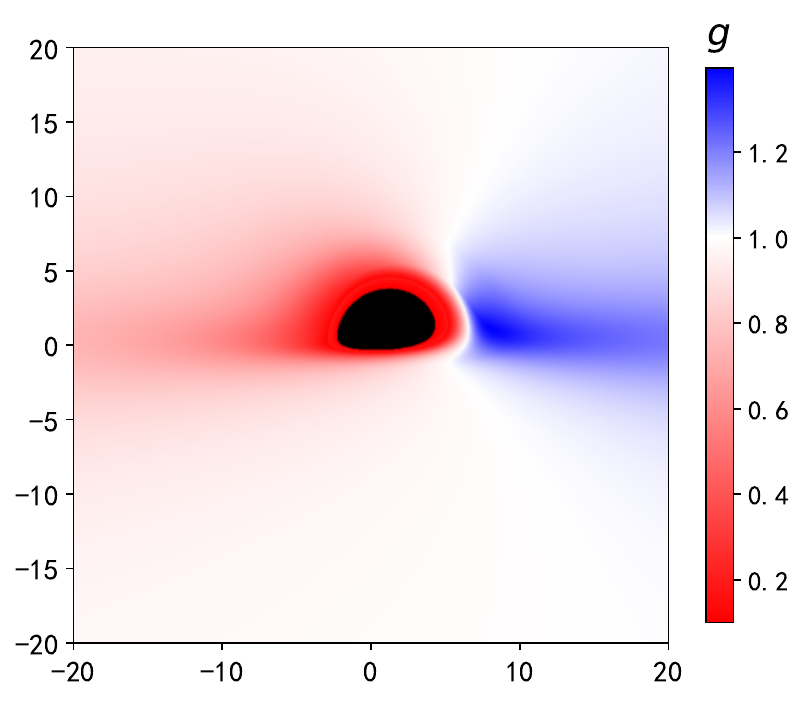}
						\put(15,85){\color{black} $a=0.95, \theta_{\mathrm{obs}}=80^{\circ}$}
					\end{overpic}
				}
			\end{minipage}
			\vspace{20pt} 
			\\ 
			\begin{minipage}[t]{0.24\textwidth}
				\centering
				\hbox{
					\begin{overpic}[width=1.0\textwidth]{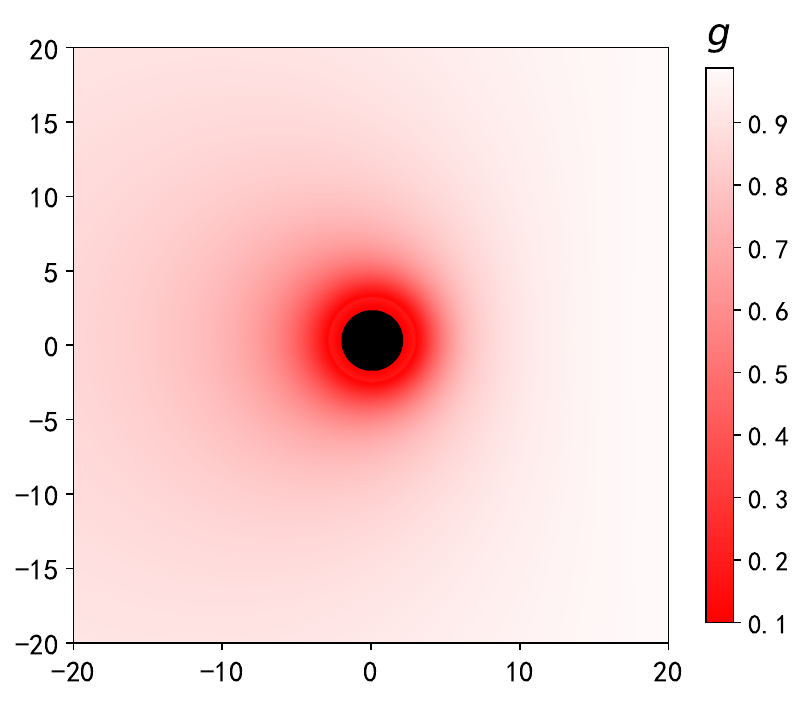}
						\put(3,85){\color{black} $a=0.5, \zeta=0.5, \theta_{\mathrm{obs}}=17^{\circ}$}
					\end{overpic}
				}
			\end{minipage}
			&
			\begin{minipage}[t]{0.24\textwidth}
				\centering
				\hbox{
					\begin{overpic}[width=1.0\textwidth]{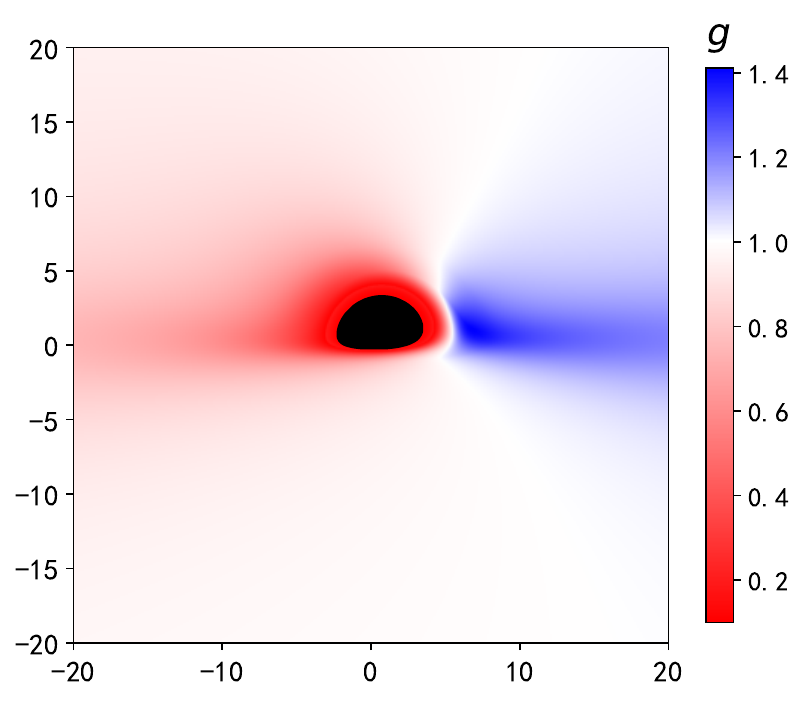}
						\put(3,85){\color{black} $a=0.5, \zeta=0.5, \theta_{\mathrm{obs}}=80^{\circ}$}
					\end{overpic}
				}
			\end{minipage}
			&
			\begin{minipage}[t]{0.24\textwidth}
				\centering
				\hbox{
					\begin{overpic}[width=1.0\textwidth]{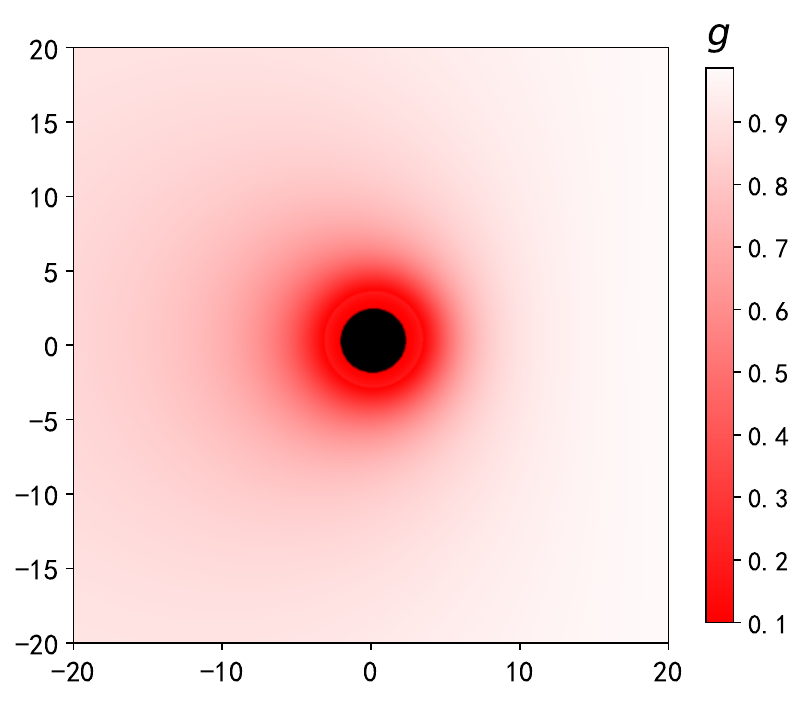}
						\put(1,85){\color{black} $a=0.95, \zeta=0.15, \theta_{\mathrm{obs}}=17^{\circ}$}
					\end{overpic}
				}
			\end{minipage}
			&
			\begin{minipage}[t]{0.24\textwidth}
				\centering
				\hbox{
					\begin{overpic}[width=1.0\textwidth]{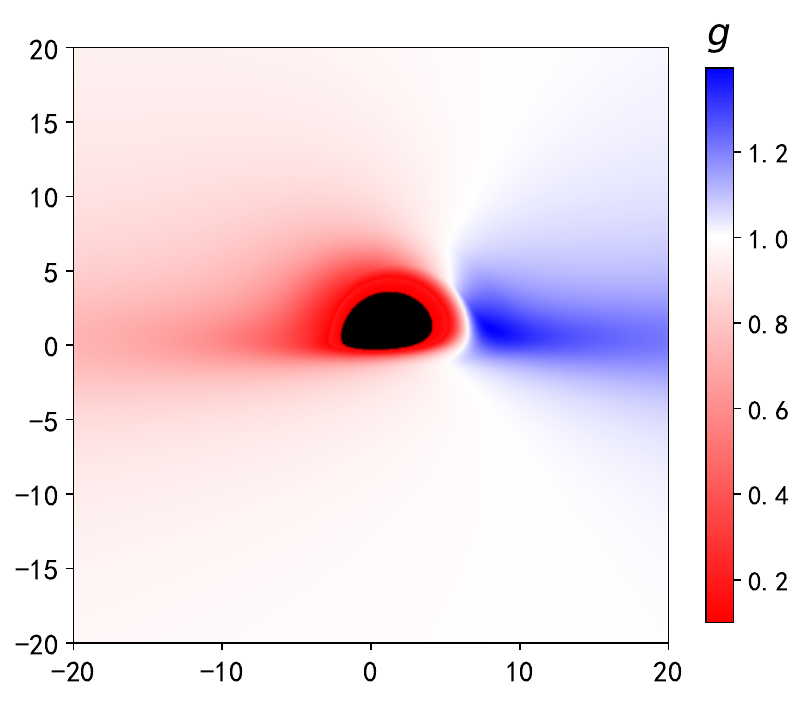}
						\put(1,85){\color{black} $a=0.95, \zeta=0.15, \theta_{\mathrm{obs}}=80^{\circ}$}
					\end{overpic}
				}
			\end{minipage}
		\end{tabular}
		\caption{The redshift distribution of the first-order (direct) image from retrograde thin accretion disks, illustrating the dependence on spacetime parameters and viewing angle. The upper panel corresponds to a Kerr black hole, while the lower panel shows the rotating Ay\'on-Beato--Garc\'{\i}a solution.}
		\label{hongyin0}
	\end{figure*}

\begin{figure*}[htbp]
		\centering
		\setlength{\tabcolsep}{1pt} 
		\begin{tabular}{cccc}
			\begin{minipage}[t]{0.24\textwidth}
				\centering
				\hbox{
					\begin{overpic}[width=1.0\textwidth]{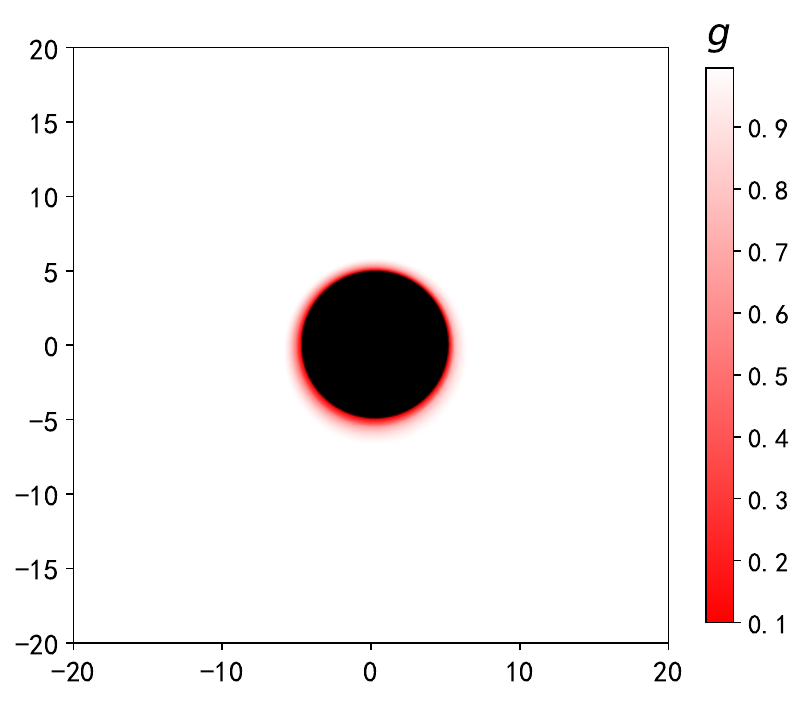}
						\put(15,85){\color{black} $a=0.5, \theta_{\mathrm{obs}}=17^{\circ}$} 
					\end{overpic}
				}
			\end{minipage}
			&
			\begin{minipage}[t]{0.24\textwidth}
				\centering
				\hbox{
					\begin{overpic}[width=1.0\textwidth]{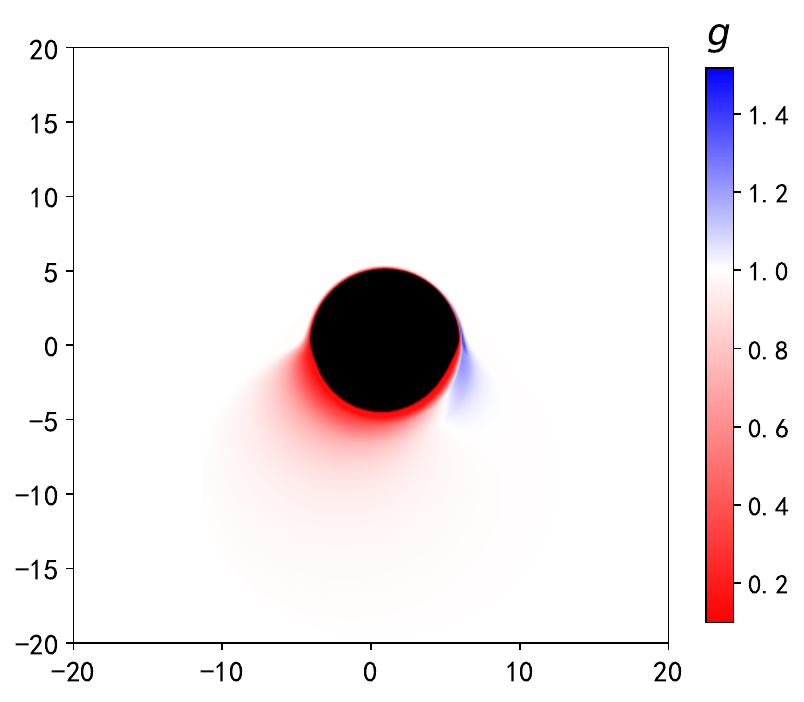}
						\put(15,85){\color{black} $a=0.5, \theta_{\mathrm{obs}}=80^{\circ}$}
					\end{overpic}
				}			
			\end{minipage}
			&
			\begin{minipage}[t]{0.24\textwidth}
				\centering
				\hbox{
					\begin{overpic}[width=1.0\textwidth]{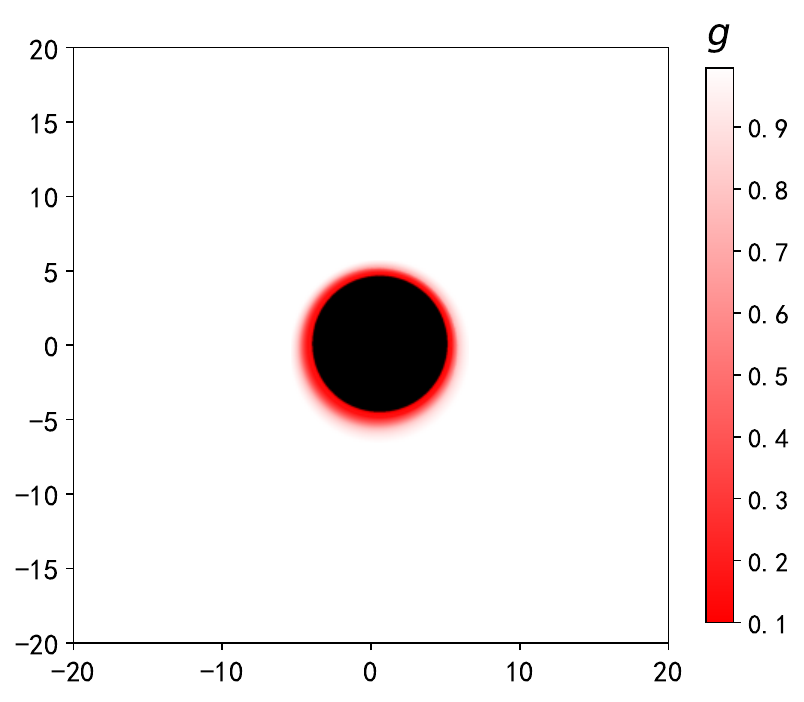}
						\put(15,85){\color{black} $a=0.95, \theta_{\mathrm{obs}}=17^{\circ}$}
					\end{overpic}
				}
			\end{minipage}
			&
			\begin{minipage}[t]{0.24\textwidth}
				\centering
				\hbox{
					\begin{overpic}[width=1.0\textwidth]{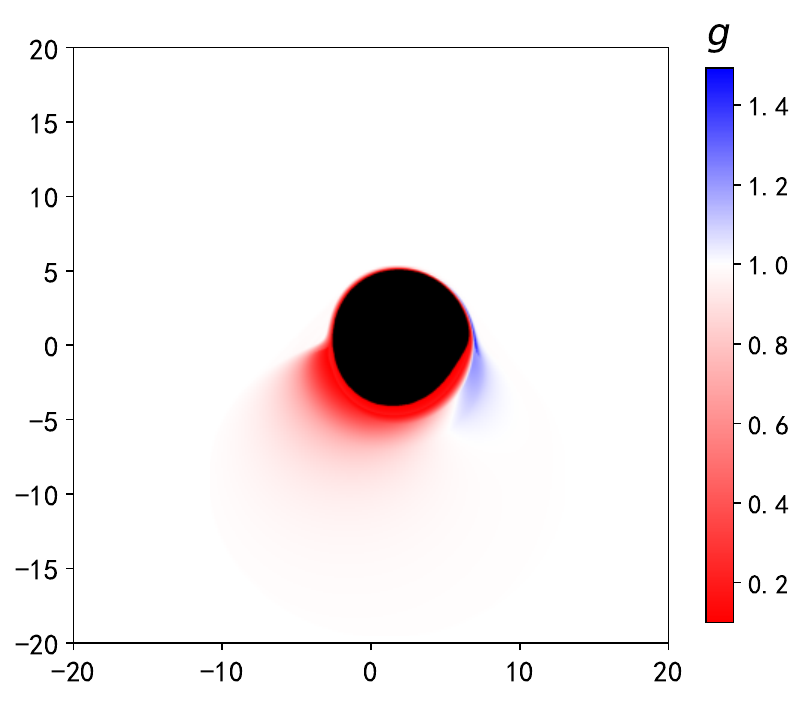}
						\put(15,85){\color{black} $a=0.95, \theta_{\mathrm{obs}}=80^{\circ}$}
					\end{overpic}
				}
			\end{minipage}
			\vspace{20pt} 
			\\ 
			\begin{minipage}[t]{0.24\textwidth}
				\centering
				\hbox{
					\begin{overpic}[width=1.0\textwidth]{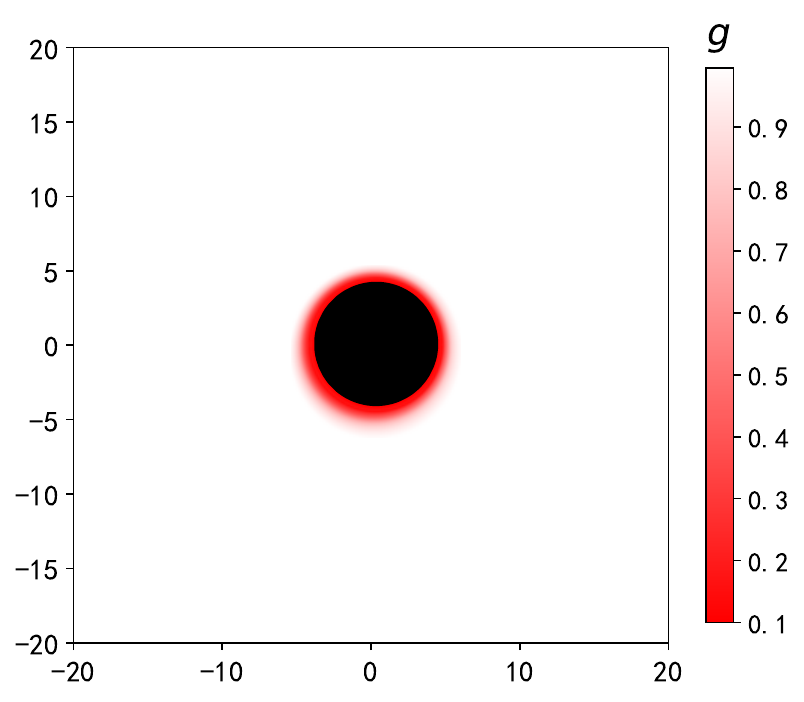}
						\put(3,85){\color{black} $a=0.5, \zeta=0.5, \theta_{\mathrm{obs}}=17^{\circ}$}
					\end{overpic}
				}
			\end{minipage}
			&
			\begin{minipage}[t]{0.24\textwidth}
				\centering
				\hbox{
					\begin{overpic}[width=1.0\textwidth]{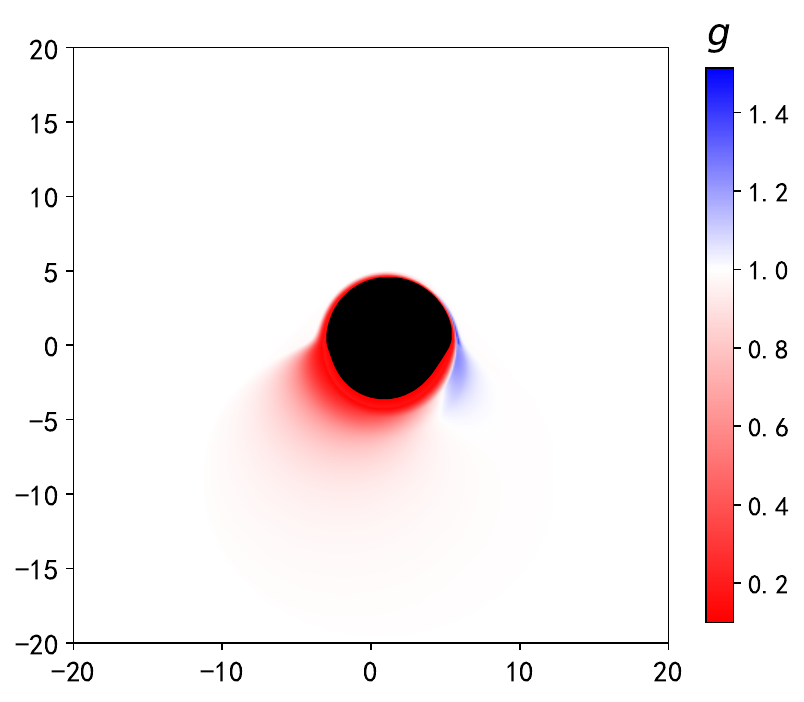}
						\put(3,85){\color{black} $a=0.5, \zeta=0.5, \theta_{\mathrm{obs}}=80^{\circ}$}
					\end{overpic}
				}
			\end{minipage}
			&
			\begin{minipage}[t]{0.24\textwidth}
				\centering
				\hbox{
					\begin{overpic}[width=1.0\textwidth]{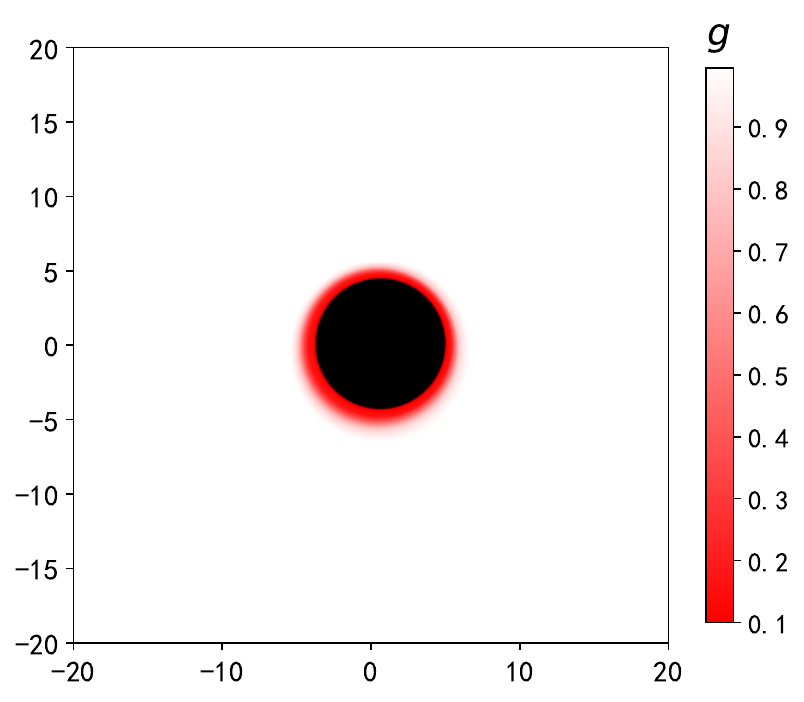}
						\put(1,85){\color{black} $a=0.95, \zeta=0.15, \theta_{\mathrm{obs}}=17^{\circ}$}
					\end{overpic}
				}
			\end{minipage}
			&
			\begin{minipage}[t]{0.24\textwidth}
				\centering
				\hbox{
					\begin{overpic}[width=1.0\textwidth]{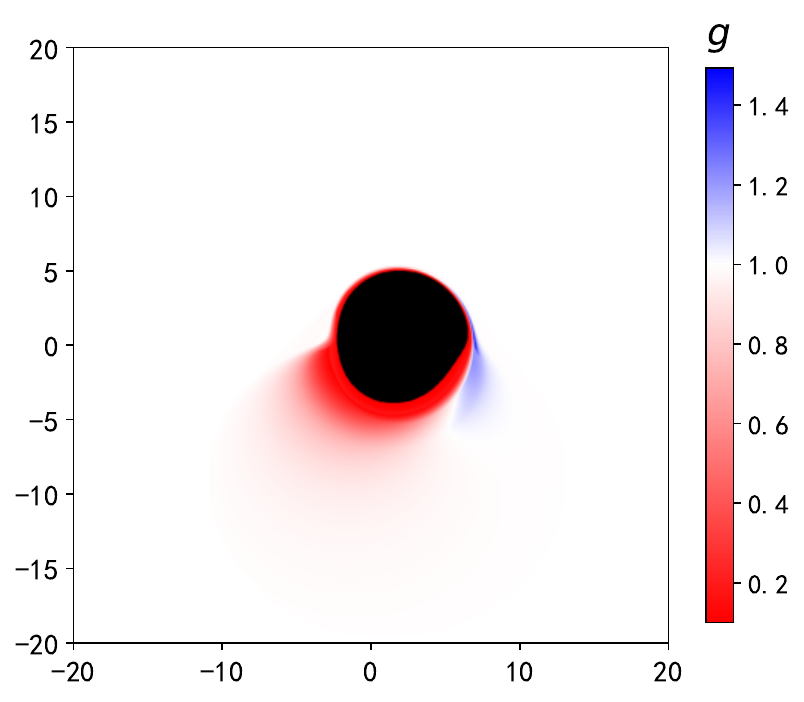}
						\put(1,85){\color{black} $a=0.95, \zeta=0.15, \theta_{\mathrm{obs}}=80^{\circ}$}
					\end{overpic}
				}
			\end{minipage}
		\end{tabular}
		\caption{The redshift distribution of the second-order (lensed) image from retrograde thin accretion disks, illustrating the dependence on spacetime parameters and viewing angle. The upper panel corresponds to a Kerr black hole, while the lower panel shows the rotating Ay\'on-Beato--Garc\'{\i}a solution.}
		\label{hongyin1}
	\end{figure*}

We now summarize the numerical setup used for all images and redshift maps. The observer is placed at a large radius $r_{\mathrm{o}}$ in the asymptotically flat region; numerically, $r_{\mathrm{o}}$ is chosen large enough that the results are insensitive to its precise value, so that the observer is effectively at infinity. The screen coordinates $(x, y)$ are measured in units of $M$, and each image covers the field of view $x, y \in [-20M, 20M]$. The image plane is sampled on a base grid of spacing $0.1M$, adaptively refined down to $M/35$ near the critical curve, where the transfer functions vary rapidly; we have checked that further refinement does not visibly change the images. For each pixel, the backward-integrated null geodesic is followed and the first two equatorial crossings are recorded, so that the displayed intensity contains the direct ($n = 1$) and first-order lensed ($n = 2$) images; higher-order images are exponentially demagnified and contribute negligibly at this resolution. The weighting factors are set to $f_{\mathrm{n}} = 1$, so the intensity $I_{\nu_{\mathrm{o}}}$ in Figs. \ref{xijis} and \ref{xijin} is given in arbitrary units, with each panel normalized independently; the color bars therefore indicate relative brightness within each panel and carry no absolute calibration. The redshift factor $g$ shown in Figs. \ref{hongyis0}--\ref{hongyin1} is dimensionless.

Figures \ref{xijis} and \ref{xijin} display simulated images of a Kerr black hole (top panels) and a rotating Ay\'on-Beato--Garc\'{\i}a (ABG) black hole (bottom panels), both illuminated by an optically thin accretion disk. The disk is modeled with prograde flow in Fig. \ref{xijis} and retrograde flow in Fig. \ref{xijin}. The observer is located at the angle of inclination $\theta_{\mathrm{obs}} = 17^\circ$ and $\theta_{\mathrm{obs}} = 80^\circ$, respectively. Irrespective of whether the accretion flow is prograde or retrograde, the separation between direct emission and higher-order lensed images becomes increasingly pronounced with increasing observer inclination $\theta_{\mathrm{obs}}$.

	Figures \ref{hongyis0}-\ref{hongyin1} present the redshift distribution maps for both direct and lensed emission from the accretion disk, covering prograde and retrograde flow configurations. As the observer’s inclination angle $\theta_{\mathrm{obs}}$ increases from $17^\circ$ to $80^\circ$, a systematic trend emerges: the redshifted region progressively contracts, while the blueshifted region expands correspondingly. This behavior reflects the growing dominance of Doppler boosting from the approaching side of the disk at higher inclinations.
	
	\section{EHT Constraints on Black Hole Parameters from M87$^{*}$ and Sgr A$^{*}$}\label{section5}
	
	Significant constraints on theoretical black hole models are imposed by the Event Horizon Telescope observations of M87$^{*}$ and Sgr A$^{*}$. The reconstructed images of their shadows and photon rings allow the parameter space of non-Kerr geometries to be systematically restricted. Here, the Event Horizon Telescope (EHT) data are employed to place observational bounds on the rotating Ay\'on-Beato--Garc\'{\i}a black hole solution.

We begin by emphasizing the assumptions and limitations of this comparison. First, in accordance with the purely geometric treatment of photon propagation adopted in Sec.~\ref{section2}, the quantity we compute is the geometric shadow diameter of the background metric; possible NLED effective-metric corrections to the propagation of the actual electromagnetic radiation are not included. Second, the EHT does not observe a pristine mathematical shadow but a bright emission ring, and the mapping between the measured ring diameter and the theoretical shadow diameter is model dependent; we therefore adopt the approach widely used in similar studies and compare the geometric shadow diameter directly with the calibrated angular-size intervals quoted by the EHT collaboration. Third, we fix the mass, distance, and inclination of each source to representative central values and do not propagate their uncertainties, nor those of the emission model and calibration, through a likelihood analysis. For these reasons the parameter ranges quoted below should be regarded as indicative estimates rather than statistically rigorous bounds, and we round all quoted values to two decimal places, which is more than sufficient given the systematic uncertainties just described.

	We clarify the apparent tension with Ramadhan et al. \cite{Ramadhan:2023ogm}, who found that the ABG black hole shadow is inconsistent with Sgr A$^*$ observations. Their work employs a static ABG metric with an extra NLED effective geometric correction that strongly reduces the shadow size. In contrast, our rotating ABG metric (and its static limit) corresponds to a different regular solution without such an effective geometry factor. The small shadow in their study is a unique feature of their specific model, not a generic property of ABG black holes. Our solution yields a shadow size fully compatible with EHT observations of Sgr A$^*$. The difference between the two treatments is precisely the effective-metric issue discussed in Sec.~\ref{section2}: Ref. \cite{Ramadhan:2023ogm} propagates light on an NLED effective optical geometry of a static ABG solution, whereas we compute the purely geometric shadow of the rotating metric (\ref{xianyuan}). The strong reduction of the shadow size found there illustrates how large such corrections can in principle be, and it is one of the main reasons why we present our EHT comparison as an indicative, geometry-level estimate.
	
	\subsection{Constraining Black Hole Parameters with M87$^{*}$ EHT Data}
	The black hole shadow is characterized by an areal radius $R_\mathrm{s} = \sqrt{A/\pi}$, where $A$ denotes the shadow area. As observed from Earth, this corresponds to an angular diameter
	\begin{equation}
		\theta_\mathrm{d} = 2\,\frac{R_\mathrm{s}}{d} = \frac{2}{d}\sqrt{\frac{A}{\pi}},
		\label{seid}
	\end{equation}
	with $d$ being the distance to the source. For M87$^{*}$, we adopt $M = 6.5 \times 10^{9}\,M_\odot$ and $d = 16.8\,\mathrm{Mpc}$, again fixing both quantities to their central values. 
	
\begin{figure*}[htbp]
		\begin{minipage}[t]{0.45\textwidth}
			\centering
			\begin{overpic}[width=0.8\textwidth]{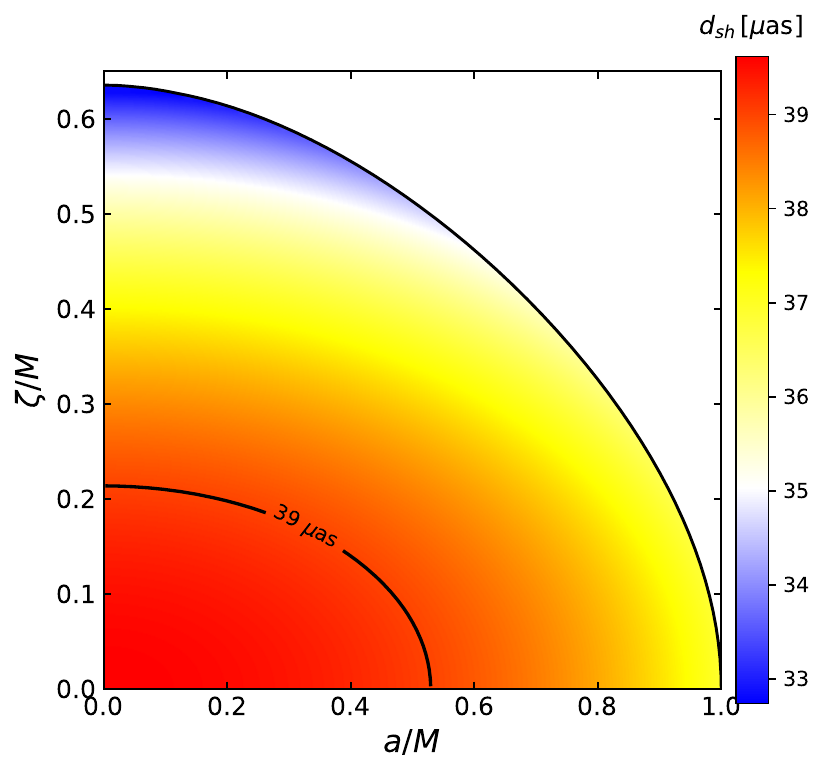} 
				\put(45,90){\color{black}\large $\theta_{\mathrm{obs}}=17^\circ$} 
			\end{overpic}
		\end{minipage}
		\hfill 
		\begin{minipage}[t]{0.45\textwidth}
			\centering
			\begin{overpic}[width=0.8\textwidth]{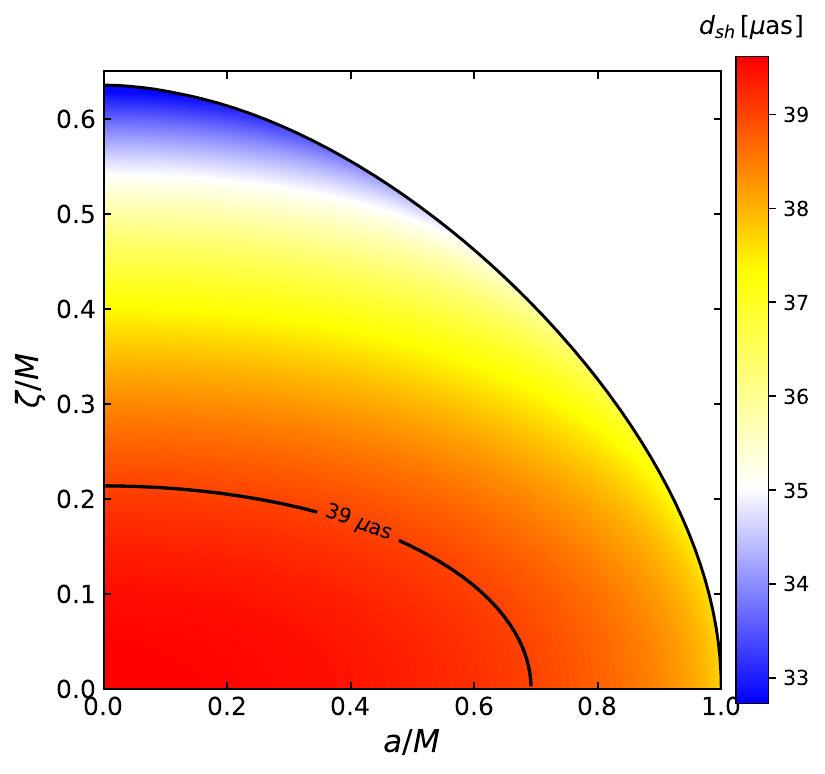} 
				\put(45,90){\color{black}\large $\theta_{\mathrm{obs}}=90^\circ$}
			\end{overpic}
		\end{minipage}
		\caption{Density plots of $\theta_d$ for different $\theta_{\mathrm{obs}}$: Left, $\theta_{\mathrm{obs}} = 17^\circ$; Right, $\theta_{\mathrm{obs}} = 90^\circ$. Observed M87$^{*}$ diameter: $\theta_d = 42 \pm 3\,\mu\mathrm{as}$. Black line: $\theta_d = 39\,\mu\mathrm{as}$. The upper limit $\theta_d = 45\,\mu\mathrm{as}$ is satisfied everywhere in the displayed parameter space, so only the $39\,\mu\mathrm{as}$ contour appears; the region above the black line is excluded.}
		\label{M87}
	\end{figure*}

	Figure \ref{M87} displays the dependence of $\theta_\mathrm{d}$ on the spin parameter $a$ and charge parameter $\zeta$ for rotating Ay\'on-Beato--Garc\'{\i}a black holes, evaluated at two representative inclination angles: $\theta_{\mathrm{obs}} = 17^\circ$ (near face-on) and $\theta_{\mathrm{obs}} = 90^\circ$ (edge-on). The angle $17^\circ$ corresponds to the inclination of the large-scale jet of M87, which is believed to be aligned with the black hole spin axis and was adopted by the EHT collaboration in its own analyses \cite{EventHorizonTelescope:2019dse}; the edge-on case $90^\circ$ is included as a limiting configuration that maximizes the asymmetry of the shadow induced by the spin. The Event Horizon Telescope measurement constrains the shadow diameter of M87$^{*}$ to the $1\sigma$ interval $39\,\mu\mathrm{as} \leq \theta_\mathrm{d} \leq 45\,\mu\mathrm{as}$. Adopting the central values of the mass and distance of M87$^{*}$ and imposing this interval on the geometric shadow diameter yields the following indicative estimates, rounded to two decimal places:
	
	\begin{itemize}
		\item For $\theta_{\mathrm{obs}} = 17^\circ$: \quad $a \lesssim 0.53\,M$, \quad $\zeta \lesssim 0.21\,M$;
		\item For $\theta_{\mathrm{obs}} = 90^\circ$: \quad $a \lesssim 0.69\,M$, \quad $\zeta \lesssim 0.21\,M$.
	\end{itemize}
	
	These results demonstrate that the allowed parameter space depends sensitively on the viewing geometry. We stress that the quoted intervals are the projections of a two-dimensional allowed region in the $(a, \zeta)$ plane onto the two axes: the allowed values of $a$ and $\zeta$ are strongly correlated, and the upper bound on $\zeta$ is attained only for slowly spinning configurations, as can be seen in Fig. \ref{M87}. The Kerr limit $\zeta = 0$ is fully consistent with the M87$^{*}$ measurement for all inclinations considered.
	
	\subsection{Constraining Black Hole Parameters with Sgr A$^{*}$ EHT Data}
	The EHT collaboration released the first resolved image of the shadow of Sgr A$^{*}$ based on the 2017 VLBI campaign at 1.3\,mm wavelength. This observation is particularly valuable for probing strong-field gravity: (i) Sgr A$^{*}$ samples spacetime curvature $\sim 10^6$ times stronger than that of M87$^{*}$, and (ii) its mass-to-distance ratio is independently constrained by decades of stellar orbit monitoring. 
	
	Using multiple imaging and modeling pipelines, the EHT data reveal a remarkably consistent ring-like structure. Independent measurements of the S0-2 star’s orbit with the Keck telescopes and the Very Large Telescope Interferometer (VLTI) yield a mass of $M = 4.0_{-0.6}^{+1.1} \times 10^6\,M_\odot$ and a distance of $D = 8$\,kpc. The reconstructed shadow diameter falls within the range $\theta_\mathrm{d} \in [46.9,\,50]\,\mu\mathrm{as}$, in excellent agreement with the Kerr black hole prediction. Nevertheless, this measurement still leaves room for viable alternatives within modified theories of gravity. In the following analysis we fix the mass and distance of Sgr A$^{*}$ to their central values, $M = 3.93 \times 10^{6}\,M_\odot$ and $D = 8\,\mathrm{kpc}$. Both quantities carry observational uncertainties of about $10\%$, and these uncertainties are not folded into our analysis. Consequently, the parameter ranges derived below depend directly on the adopted central values; the quantitative consequence of this choice is made explicit at the end of this subsection.
	
\begin{figure*}[htbp]
		\begin{minipage}[t]{0.45\textwidth}
			\centering
			\begin{overpic}[width=0.8\textwidth]{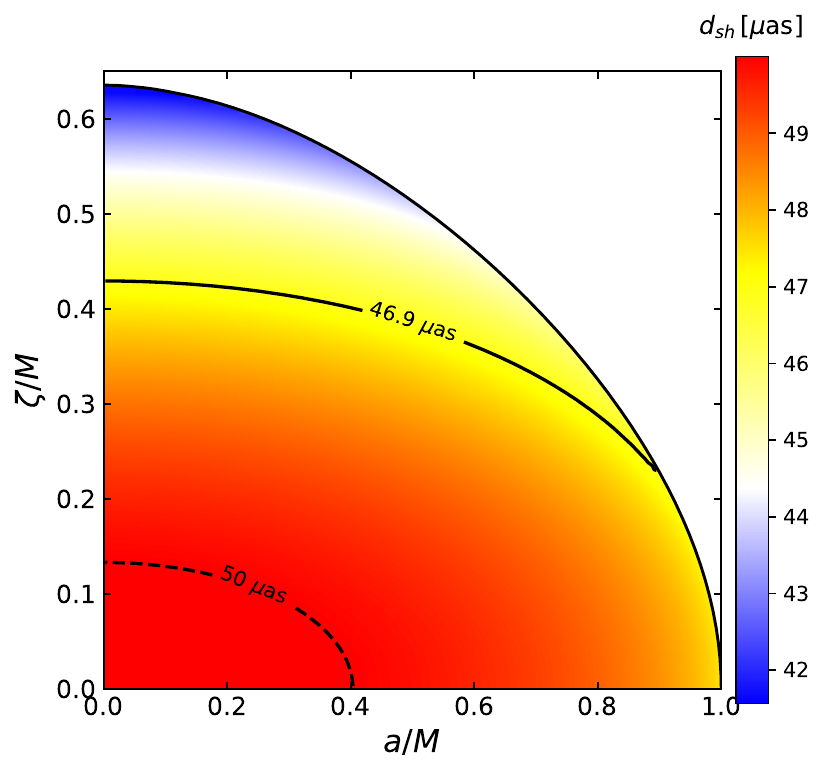} 
				\put(45,90){\color{black}\large $\theta_{\mathrm{obs}}=50^\circ$} 
			\end{overpic}
		\end{minipage}
		\hfill 
		\begin{minipage}[t]{0.45\textwidth}
			\centering
			\begin{overpic}[width=0.8\textwidth]{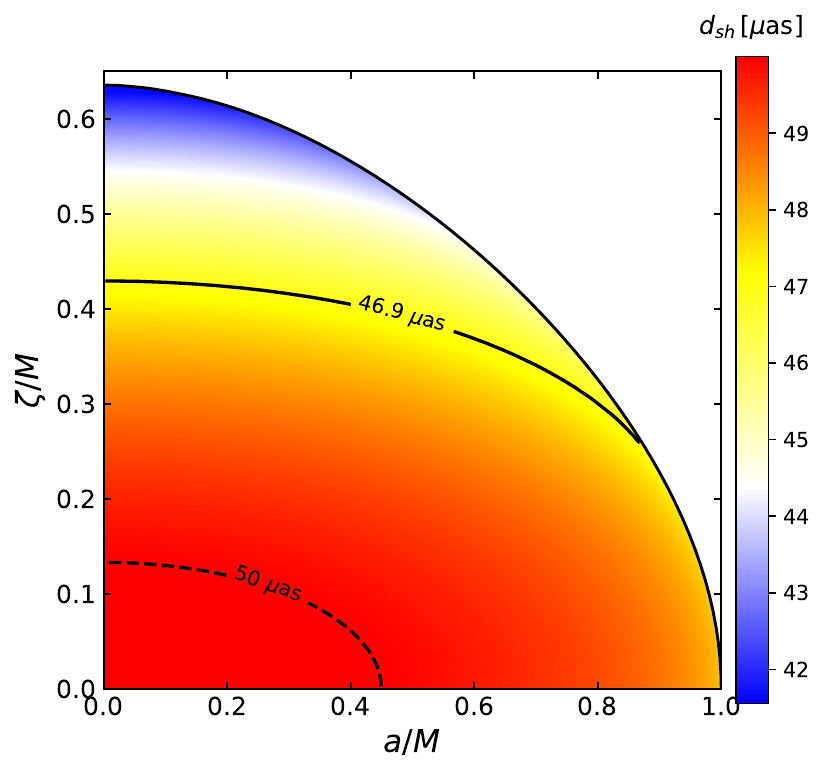} 
				\put(45,90){\color{black}\large $\theta_{\mathrm{obs}}=90^\circ$}
			\end{overpic}
		\end{minipage}
		\caption{Density plots of $\theta_d$ for different $\theta_{\mathrm{obs}}$: Left, $\theta_{\mathrm{obs}} = 50^\circ$; Right, $\theta_{\mathrm{obs}} = 90^\circ$. Observed Sgr A$^{*}$ diameter: $\theta_d \in [46.9,\,50]\,\mu\mathrm{as}$. Solid black line: $\theta_d = 46.9\,\mu\mathrm{as}$; Dashed black line: $\theta_d = 50\,\mu\mathrm{as}$.}
		\label{SgrA}
	\end{figure*}

	Figure \ref{SgrA} shows the predicted angular shadow diameter $\theta_\mathrm{d}$ for rotating Ay\'on-Beato--Garc\'{\i}a black holes as a function of the spin parameter $a$ and charge parameter $\zeta$, evaluated at two representative inclinations: $\theta_{\mathrm{obs}} = 50^\circ$ and $\theta_{\mathrm{obs}} = 90^\circ$. The EHT analysis of Sgr A$^{*}$ disfavors large inclinations and prefers viewing angles $\theta_{\mathrm{obs}} \lesssim 50^\circ$ \cite{EventHorizonTelescope:2022urf}; we therefore adopt $50^\circ$ as a representative upper value of the preferred range, and again include the edge-on case $90^\circ$ as a limiting configuration. The observed bounds for Sgr A$^{*}$ are indicated by vertical lines: the solid black line marks $\theta_\mathrm{d} = 46.9\,\mu\mathrm{as}$ (lower limit), and the dashed black line denotes $\theta_\mathrm{d} = 50\,\mu\mathrm{as}$ (upper limit).
	
	With the adopted central values of $M$ and $D$, the allowed region in the $(a, \zeta)$ plane is the band between the two contours in Fig. \ref{SgrA}. Since $a$ and $\zeta$ are strongly correlated along this band, quoting independent intervals for the two parameters would be misleading; instead we describe the band by its intersections with the coordinate axes, with all values rounded to two decimal places:
	\begin{itemize}
		\item For $\theta_{\mathrm{obs}} = 50^\circ$: along the $\zeta = 0$ axis, which corresponds to the Kerr limit, the allowed spins are $0.40 \lesssim a \lesssim 0.90$, while along the $a = 0$ axis the allowed charges are $0.13\,M \lesssim \zeta \lesssim 0.43\,M$;
		\item For $\theta_{\mathrm{obs}} = 90^\circ$: along $\zeta = 0$ the allowed spins are $0.45 \lesssim a \lesssim 0.86$, with the same intersection $0.13\,M \lesssim \zeta \lesssim 0.43\,M$ along $a = 0$.
	\end{itemize}
	In particular, the overall upper bound from Sgr A$^{*}$ is $\zeta \lesssim 0.43\,M$, attained in the non-rotating limit.

	The appearance of a nonvanishing lower edge of the allowed band calls for a careful explanation, since the EHT results are routinely described as consistent with the Kerr geometry. This feature is entirely an artifact of fixing the mass and distance to the central values adopted above: with $M = 3.93 \times 10^{6}\,M_\odot$ and $D = 8\,\mathrm{kpc}$, the predicted shadow diameter of a Schwarzschild black hole is $\theta_\mathrm{d} \simeq 50.4\,\mu\mathrm{as}$, which exceeds the adopted upper limit of $50\,\mu\mathrm{as}$ by less than $1\%$; a slight reduction of the shadow size, provided either by spin or by the charge parameter, is then formally required. This mild tension disappears entirely once the uncertainties of about $10\%$ in the mass-to-distance ratio of Sgr A$^{*}$ are taken into account, and it must not be interpreted as observational evidence either for a nonzero $\zeta$ or against the Kerr geometry. This example illustrates concretely why the ranges quoted in this section are indicative estimates rather than statistical bounds.
	
	Finally, we comment on the relation between the estimates obtained from the two sources. Unlike a universal coupling constant of a modified theory of gravity, the charge parameter $\zeta$ is a property of each individual black hole, in the same way as its mass and spin: there is no a priori reason for M87$^{*}$ and Sgr A$^{*}$ to carry the same value of $\zeta$ in units of their respective masses. Intersecting the two allowed regions would therefore not be meaningful, and we do not perform a combined analysis; the estimates $\zeta \lesssim 0.21\,M$ for M87$^{*}$ and $\zeta \lesssim 0.43\,M$ for Sgr A$^{*}$ are quoted separately for each source.

	\section{Conclusion} 
	
	\label{section6}
	We have systematically investigated the shadow geometry, accretion disk structure, and synthetic observational images of rotating Ay\'on-Beato--Garc\'{\i}a (ABG) black holes, characterized by mass $M$, spin $a$, and the NLED charge parameter $\zeta$. Throughout, photons were treated as neutral test particles following null geodesics of the background metric, so all quantities related to the shadow refer to the purely geometric shadow of the spacetime (Sec.~\ref{section2}). Our analysis demonstrates that the shadow size exhibits a monotonic decrease with increasing $\zeta$, reflecting the influence of the regularizing charge on spacetime curvature. Notably, in near-extremal spin regimes (e.g., $a = 0.95$), the shadow develops a pronounced ``D''-shaped asymmetry—a distinctive signature absent in the Kerr case—arising from the interplay between frame-dragging and the modified photon orbit structure induced by $\zeta$.
	
	The thermal properties of the disk, including the energy flux, the temperature, and the spectral energy distribution, were computed within the standard Novikov--Thorne model, whose inner edge lies at the ISCO; consistently with Kerr disks, both the NLED charge parameter $\zeta$ and the prograde spin $a$ move the ISCO inward, raise the radiative efficiency, and enhance the peak flux and temperature. For the synthetic images, by contrast, we adopted a separate, phenomenological optically thin emission model in which the emitting region extends down to the event horizon, with circular geodesic motion outside the ISCO and critical plunging orbits inside it. This treatment reveals that the combined dependence of image morphology on $(a, \zeta)$ and the observer’s inclination angle $\theta_{\mathrm{obs}}$ plays a crucial role in shaping both the overall brightness asymmetry and the fine structure of the inner shadow. At high inclinations ($\theta_{\mathrm{obs}} \gtrsim 50^\circ$), the direct and higher-order lensed images become spatially resolved, giving rise to a characteristic hat-like feature in the total intensity map.
	
	Furthermore, we compute the redshift distribution of both direct and lensed photons across a range of parameters and viewing angles, providing a detailed map of Doppler and gravitational shifts imprinted by the ABG spacetime. Finally, we compared the geometric shadow diameters with the EHT measurements of M87$^{*}$ and Sgr A$^{*}$ at representative inclinations of $17^\circ$, $50^\circ$, and $90^\circ$, adopting the central values of the measured masses and distances. This comparison yields the indicative estimates $\zeta \lesssim 0.21\,M$ from M87$^{*}$ and $\zeta \lesssim 0.43\,M$ from Sgr A$^{*}$, with a strong correlation between the allowed values of $a$ and $\zeta$ within each source. Since $\zeta$ characterizes each black hole individually rather than a universal coupling of the theory, the two estimates are quoted separately and are not combined. These numbers do not include the systematic uncertainties inherent in the comparison, namely the geometric treatment of photon propagation, the model-dependent mapping between the observed emission ring and the theoretical shadow, and the uncertainties in the masses, distances, and inclinations of the sources; the Kerr limit $\zeta = 0$ remains fully consistent with both observations. In particular, the apparent lower bound on $\zeta$ or on $a$ suggested by the Sgr A$^{*}$ comparison is an artifact of fixing the mass and distance to their central values and carries no statistical significance.

	These results illustrate both the phenomenological richness of rotating regular black hole spacetimes and the current limitations of shadow-based tests. Several extensions of this work suggest themselves: a full likelihood analysis propagating the mass, distance, inclination, and calibration uncertainties of the EHT measurements; the construction of the NLED effective optical geometry to quantify the corrections to the purely geometric shadow; and an assessment of the dynamical stability of the rotating ABG spacetime in light of the instabilities recently identified for nonsingular NLED black holes \cite{DeFelice:2024seq}. High-resolution VLBI imaging, together with such refinements, will progressively sharpen the extent to which regular black hole models can be distinguished from the Kerr paradigm.
	
	\begin{acknowledgments}
		This study is supported in part by National Natural Science Foundation of China (Grant
		No. 12333008).
	\end{acknowledgments}

\end{document}